\documentclass[showpacs,showkeys, preprintnumbers,pre,amsmath,amssymb,superscriptaddress]{revtex4}
\usepackage{graphicx}
\usepackage{booktabs}
\usepackage{array}
\usepackage{hyperref}

\usepackage{bm}
\usepackage{times}

\usepackage{balance}

\usepackage{epic,eepic}
\usepackage{graphics}
\usepackage{epsfig}
\usepackage{subfigure}

\newcommand{\de}{\partial}

\newcommand{\al}{\alpha}

\usepackage[latin1]{inputenc}
\usepackage{pstricks,pst-node,pst-text,pst-3d}
\usepackage{amsmath}
\usepackage{graphicx}
\usepackage{pst-grad}
\usepackage{pst-vue3d}
\newpsstyle{GradGrayWhite}{fillstyle=gradient,%
    gradbegin=blue,gradend=white,linewidth=0.1mm}%
\usepackage{graphicx}
\usepackage{pst-grad}
\usepackage{pst-vue3d}
\usepackage{soul}

\newcommand{\be}{\begin{equation}}
\newcommand{\ee}{\end{equation}}
\newcommand{\bea}{\begin{eqnarray}}
\newcommand{\eea}{\end{eqnarray}}

\usepackage{amsmath}
\usepackage{amsfonts}
\usepackage{amssymb}
\usepackage{bm}
\usepackage{color}
\usepackage{graphicx}
\begin{document}

\title{Sliding drops across alternating hydrophobic and hydrophilic stripes}

\author{M. Sbragaglia}\email[]{sbragaglia@roma2.infn.it}
\affiliation{Department of Physics and INFN, University of ``Tor Vergata'', Via della Ricerca Scientifica 1, 00133 Rome, Italy\\}
\author{L. Biferale}
\affiliation{Department of Physics and INFN, University of ``Tor Vergata'', Via della Ricerca Scientifica 1, 00133 Rome, Italy\\}
\author{G. Amati}
\affiliation{SCAI - SuperComputing Applications and Innovation Department CINECA - Via dei Tizii, 6 - 00185 Roma - Italy\\}
\author{S. Varagnolo}
\affiliation{Dipartimento di Fisica e Astronomia ``G. Galilei'' and CNISM, Universit\'{a} di Padova, Via Marzolo 8, I-35131 Padova, Italy\\}
\author{D. Ferraro}
\affiliation{Dipartimento di Fisica e Astronomia ``G. Galilei'' and CNISM, Universit\'{a} di Padova, Via Marzolo 8, I-35131 Padova, Italy\\}
\author{G. Mistura}
\affiliation{Dipartimento di Fisica e Astronomia ``G. Galilei'' and CNISM, Universit\'{a} di Padova, Via Marzolo 8, I-35131 Padova, Italy\\}
\author{M. Pierno}\email[]{matteo.pierno@unipd.it}
\affiliation{Dipartimento di Fisica e Astronomia ``G. Galilei'' and CNISM, Universit\'{a} di Padova, Via Marzolo 8, I-35131 Padova, Italy\\}

\begin{abstract}
We perform a joint numerical and experimental study to \textcolor{black}{systematically} characterize the motion of \textcolor{black}{30 $\mu\textrm{l}$ drops of \textcolor{black}{pure} water and \textcolor{black}{of} ethanol in water \textcolor{black}{solutions},} sliding over a periodic array of alternating hydrophobic and hydrophilic stripes with large wettability contrast and typical width of hundreds of \textcolor{black}{microns}. The fraction of the hydrophobic \textcolor{black}{areas} has been varied from about $20$\% to $80$\%. The effects of the heterogeneous patterning can be described by a renormalized value of the critical Bond number, i.e. the critical dimensionless force needed to depin the drop before it starts to move. Close to the critical Bond number we observe a \textcolor{black}{jerky} motion characterized by an evident stick-slip dynamics. As a result, dissipation is strongly localized in time, and the mean velocity of the drops can easily decrease by an order of magnitude compared to the sliding on homogeneous surface. Lattice Boltzmann (LB) numerical simulations are crucial for disclosing to what extent the sliding dynamics can be deduced from the computed balance of capillary, viscous and body forces at varying the Bond number, the surface composition and the liquid viscosity. Away from the critical Bond number, we characterize both experimentally and numerically the dissipation inside the droplet by studying the relation between the average velocity and the applied volume forces.
\end{abstract}



\date{\today}
\maketitle

\section{Introduction}

In the last ten years surface topography has proved to be a promising tool for controlling wettability~\cite{Quere08, Seeman12, thermocapillary, Hydro} and liquid transport~\cite{LiquidTransport}. Nevertheless many challenges must still be tackled. In particular, although it is well known that the shape of a sessile drop can be controlled by the balance between capillary forces and gravity~\cite{Furmidge62,HuHScriven71}, there is still a lack of understanding on the role played by wetting and dewetting phenomena arising from the interaction with the solid substrate~\cite{Yao13}.\\
The problem of the contact line dynamics and drop motion on structured substrates has been investigated in a number of theoretical and numerical studies~\cite{Dietrich,Beltrame09,ThieleKnobloch06a,ThieleKnobloch06b,LeopoldesBucknall05,KusumaatmajaYeomans07,Kusumaatmajaeyal06,Wangetal08,Quianetal09,Herdeetal12}. In a series of works by Thiele and coworkers~\cite{Beltrame09,ThieleKnobloch06a,ThieleKnobloch06b}, the depinning process corresponding to the loss of stability of drops moving over a heterogeneous pattern has been studied in the limit of small contact angles and small wettability contrasts, with the emergence of a {\it stick-slip} motion during which the contact line jumps from one wetting defect to another~\cite{LeopoldesBucknall05,KusumaatmajaYeomans07}. Using lattice Boltzmann (LB) numerical simulations, Kusumaatmaja and coworkers~\cite{Kusumaatmajaeyal06,KusumaatmajaYeomans07} explored the feasibility of using chemical patterning to control the size and polydispersity of micrometer sized drops: in agreement with other authors~\cite{LeopoldesBucknall05} the stick-slip motion of the contact line was recorded in the simulations. Wang and coworkers~\cite{Wangetal08} simulated the moving contact line in two-dimensional chemically patterned channels using a diffuse-interface model with the generalized Navier boundary condition: the motion of the fluid-fluid interface has been found to be modulated by the chemical pattern on the surfaces, leading to a stick-slip behaviour of the contact line. In addition molecular-dynamics simulations~\cite{Quianetal09} and the Stokes equations employing a boundary element method~\cite{Herdeetal12} have been applied to the problem. From the experimental side, the sliding of a drop on a chemically striped surface has been studied~\cite{Moritaetal05,Suzukietal08}. Morita {\it et al.}~\cite{Moritaetal05} have produced micropatterned surfaces with alternating stripes of different wettability having a width ranging from $1$ to $20$ microns. Their attention is focused on the anisotropic behavior of drops sliding in the direction parallel and orthogonal to the stripes. Suzuki {\it et al.}~\cite{Suzukietal08} realized micropatterned surfaces with alternating stripes having width of $100$ or $500$ microns and a wetttability contrast of about $10^{\circ }$. They report smooth oscillations in the advancing and receding contact angles for the $500$ microns stripes and practically constant angles for the narrow stripes. For the former pattern, fluctuations in the velocity are reported.\\
Despite such an ample amount of works dealing with drops moving on chemically patterned surfaces, a joint numerical and experimental systematic investigation of the stick-slip regime, the role of the energy balance, and the effect of the patterning at the mesoscale is still lacking. Given this state of affairs, we started a systematic and comprehensive study to explore the dynamics of drops sliding down an inclined plane (see Fig.~\ref{pianoinclinato}) consisting of a periodic array of alternating hydrophobic and hydrophilic stripes with a large wettability contrast (about $70^{\circ}$). This is a case where the usual theoretical approaches relying on a long-wavelength limit of hydrodynamics~\cite{Oron,ThieleKnobloch06a,ThieleKnobloch06b} cannot provide quantitative answers, as they restrict themselves to drops with small contact angles and small wettability contrasts. For small velocities, a jerky motion featuring an evident stick-slip dynamics is observed~\cite{prl13}. The mean sliding velocity is found to be systematically affected by the patterning details, with a slowing down that can easily reach up to an order of magnitude with respect to the corresponding homogeneous coating with the same static morphology (the same equilibrium contact angle). To investigate a more ample interval of contact angles and extend the experimental observation in~\cite{prl13}, we studied sliding drops of water and ethanol in water mixtures.  Numerical simulations performed in close synergy with the experiments are crucial for disclosing the physical mechanisms behind the sliding dynamics, elucidating the relative importance of capillary, viscous and body forces, quantities otherwise impossible to obtain in the experiments.\\
The paper is organized as follows: in Sec.~\ref{sec:1a} we describe the experimental details for realizing the heterogeneous patterns and studying the sliding drops (Sec.~\ref{sec:1b}). Numerical results are presented in Sec.~\ref{sec:2}. Conclusions follow in Sec.~\ref{sec:3}. In the Appendix (Sec.~\ref{sec:appendix}) we report the details of the LB method used.

\section{Experiments}\label{sec:1}

\begin{figure}[tbp]
\includegraphics[scale=0.5]{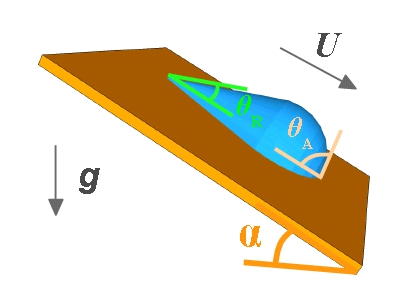}\\
\includegraphics[scale=0.5]{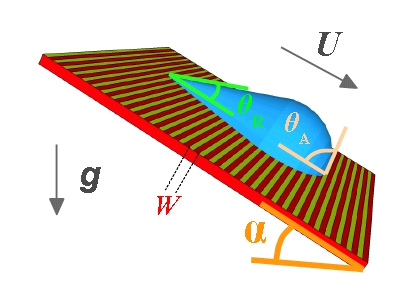}
\caption{(Color online) Sliding drop on an inclined plane tilted by an angle $\alpha$. The characteristic sliding velocity $U$ is governed by the down-plane component of the gravitational acceleration $g \sin \alpha$. The advancing contact angle $\theta_A$ is found to be larger than the receding angle $\theta_R$ ({\it contact angle hysteresis}). The surfaces may be chemically homogeneous (top panel), or functionalized with stripes of alternating wettability with periodicity $W$ (bottom panel). \label{pianoinclinato}}
\end{figure}

A liquid drop  of volume $V$ sliding down an inclined plane tilted by an angle $\alpha$ is subject to the gravity force, interfacial forces and the viscous drag. The down-plane component of the drop weight is $\rho g V \sin \alpha$, $\rho$ being the fluid density and $g$ the  gravity acceleration. The interfacial force is proportional to $\gamma_{LG} V^{1/3}\Delta_{\theta}$, where $\gamma_{LG}$ is the liquid-gas surface tension and $\Delta_{\theta}$ is a non dimensional factor depending on the contact angle distribution along the perimeter and on the perimeter shape. The viscous drag force is of the order of $c (\theta_d)\eta V^{1/3} U$, where $U$ is the drop velocity, $\eta$ is the viscosity of the liquid drop while the function $c(\theta_d)$ depends on the dynamical contact angle distribution $\theta_d$ along the perimeter of the moving droplet in contact with the surface. The function $ c(\theta_d)$ results from the viscous dissipation in the wedge and encodes the general feature that smaller contact angles are associated with higher viscous dissipation~\cite{Podgorskietal01,Kimetal02}. \textcolor{black}{Bulk dissipation is usually \textcolor{black}{smaller} than the dissipation close to the contact line~\cite{Kimetal02}}. In addition, the difference between the advancing and the receding contact angle (as shown in Fig.~\ref{pianoinclinato}) does not necessarily vanish for small velocities, a feature that is known as {\it contact angle hysteresis}. The hysteresis results in the presence of a critical angle $\alpha_c$, below which the drop is pinned~\cite{Furmidge62}. Above this threshold the force balance between gravity, viscous and capillary forces implies the following scaling law~\cite{Podgorskietal01,Kimetal02} between the {\it Capillary number} \textcolor{black}{ $\rm{Ca}=\eta$ $U/\gamma_{LG}$} and the {\it Bond number} \textcolor{black}{$\rm{Bo}= (3{\it V}/4\pi )^{2/3}\rho g \sin \alpha/\gamma_{LG}$} 
\be\label{eq:scaling}
\textcolor{black}{\rm{Ca} \propto \dfrac{\rm{Bo}-\rm{Bo_c}}{c(\theta_d)}}
\ee
where \textcolor{black}{$\rm{Bo}_c=(3{\it V}/4\pi )^{2/3}\rho g \sin \alpha_c/\gamma_{LG}$} depending on the wetting hysteresis through $\Delta_{\theta}$. It is reasonable to approximate $\theta_d \approx \theta_{eq}$, the equilibrium contact angle on the homogeneous surface, either when dynamic contact angles do not deviate severely from $\theta_{eq}$ or when the arithmetic mean of the advancing and receding contact angles is close to $\theta_{eq}$~\cite{Kimetal02}.\\ 
When drops are deposited on a surface functionalized with stripes of alternating wettability, they may assume elongated shapes, which are characterized by different contact angles in the directions perpendicular and parallel to the stripes. This morphological anisotropy has been the object of intense scrutiny in a variety of situations~\cite{Pompe00,Buehrle02,Moritaetal05,Semprebonetal09,Jansenetal12}. The equilibrium properties are well described by the Cassie-Baxter equation~\cite{cassiebaxter44}
\be\label{eq:CB}
\cos \theta_{hete}=f_1 \cos \theta_1+f_2 \cos \theta_2
\ee
which averages over the surface contact angles and $f_1$ and $f_2$ are the fractions of the surface with intrinsic equilibrium contact angle $\theta_1$ and $\theta_2$ respectively. We will take the convention to indicate with subscript `1' the more hydrophobic component. \textcolor{black}{Regardless of the anisotropy of the drop, the only important requirement is that the drop should be large enough to cover at least ten different stripes, a condition which could be reasonably assumed as representative of the whole sample composition.}

\subsection{Materials and Methods}\label{sec:1a}

\begin{figure}[tbp]
\includegraphics[scale=0.5]{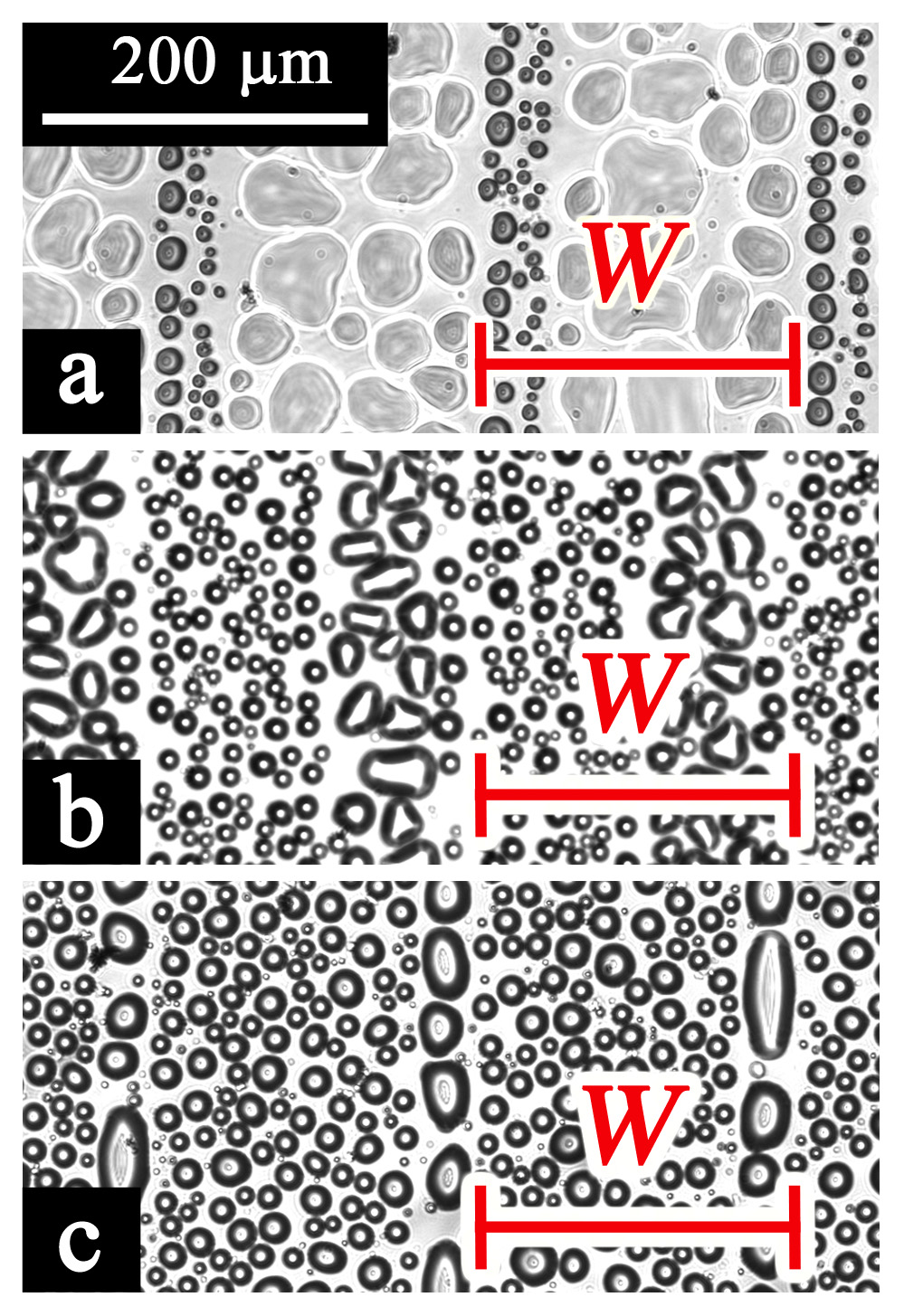}
\caption{(Color online) Vapour condensation on the heterogeneous surfaces featuring hydrophilic glass and hydrophobic OTS parallel stripes: smaller drops form on the hydrophobic areas whereas bigger drops on the hydrophilic ones. These three patterns are characterized by the same periodicity $W\sim 200\,\mu \text{m}$ (corresponding to the scale bar), but different fractions of OTS and glass: (a) 19\% OTS and 81\% glass ($f_1=0.19$, $f_2=0.81$), (b) 50\% OTS and 50\% glass ($f_1=0.5$, $f_2=0.5$) and (c) 83\% OTS and 17\% glass ($f_1=0.83$, $f_2=0.17$).  \label{condensation}}
\end{figure}

Chemically patterned surfaces, featuring alternating hydrophilic and hydrophobic stripes, are realized through microcontact printing: masters with rectangular grooves are produced by photolithography and replicated in PDMS (polydimethilsiloxane) to obtain the stamp for the printing of a solution of OTS (octadecyltrichlorosilane) in toluene on a glass substrate. The result is a surface presenting hydrophobic stripes (OTS regions) alternated with hydrophilic stripes (uncoated glass regions). Sample characterization is performed by condensing water vapour, as shown in Fig.~\ref{condensation}, where parallel stripes of different wettability can be clearly evinced having a periodicity $W\sim 200\,\mu \text{m}$.


The printed pattern is also analyzed in terms of contact angle measurements through the Cassie-Baxter equation~\cite{cassiebaxter44}, as reported in Table~\ref{angoli_cassie_baxter}.  We measured simultaneously the equilibrium contact angle both parallel ($\theta_{\parallel}$) and perpendicular ($\theta_{\bot}$) to the stripes (see cartoons in Table~\ref{angoli_cassie_baxter}) of 4 $\mu$l water drops \textcolor{black}{(which cover about 12-14 $W$)}, using the experimental apparatus described in~\cite{Tothetal11}. \textcolor{black}{The contact angle evaluation is the mean of the values measured for at least 5 independent droplets deposited on different positions on the surface and the error is their standard deviation.} In agreement with~\cite{Moritaetal05,Jansenetal12} only the equilibrium contact angle parallel to the stripes is compatible with the theoretical prediction calculated through the Cassie-Baxter equation (see Table~\ref{angoli_cassie_baxter}) and the asymmetry is more pronounced in the case of the more hydrophilic surfaces.\\
To compare the sliding of drops between heterogeneous and homogeneous surfaces, different coatings of glass slides have been produced with a variety of molecules and methods: OTS, N-octyltrimethoxysilane and Trichloro(1H,1H,2H,2H-perfluorooctyl)silane deposited from the vapour phase or by immersion in a solution of toluene, obtaining contact angles ranging from $\theta _{\textrm{eq}}=71^{\circ} \pm 2^{\circ}$ to $\theta _{\textrm{eq}}=115^{\circ} \pm 2^{\circ}$. Sliding measurements on these surfaces are performed with drops of distilled water ($\rho$ = 1000 \textcolor{black}{kg} m$^{-3}$, $\eta$ = 1 cP, $\gamma_{LG}$ = 72.8 mN m$^{-1}$ and $V$ $\approx $ 30 $\mu\textrm{l}$\textcolor{black}{, corresponding to a contact area about 30 $W$ long}) and drops of a solution of ethanol in water 30\% w/w ($\rho$ = 954 \textcolor{black}{kg} m$^{-3}$, $\eta$ = 2.5 cP, $\gamma_{LG}$ = 35.5 mN m$^{-1}$~\cite{Khattabetal12} and $V$ $\approx $ 30 $\mu\textrm{l}$\textcolor{black}{, with a length of about 30-35 $W$}) through a setup similar to~\cite{LeGrandDaerr05}. Drops of desired volume are deposited by means of a vertical syringe pump on the already inclined surface, placed on a tiltable support whose inclination angle $\alpha$ can be set with 0.1\textdegree\ accuracy. A mirror mounted under the sample holder at 45\textdegree\ with respect to the surface allows viewing the contact line and the lateral side of the drop simultaneously~\cite{LeGrandDaerr05}. The drop is lightened by two white LED backlights and is observed through a CMOS camera equipped with a macro zoom lens.  Acquired sequences of images, where drops appear dark on a light background, are analyzed through a custom made program which identifies the drop contour and then fits it with a polynomial function, subsequently used to evaluate the front and rear contact points and angles~\cite{Ferraroetal12}.

\begin{table*}[!htbp]
\caption{(Color online) Static contact angles of both homogeneous and heterogeneous surfaces of glass (red/dark) and OTS (yellow/light). \textcolor{black}{Heterogeneous samples are labeled with the corresponding OTS percentage.} In agreement with~\cite{Moritaetal05,Jansenetal12} only the static contact angle parallel to the stripes is compatible with the theoretical prediction calculated through the Cassie-Baxter equation (\ref{eq:CB}). }
\label{angoli_cassie_baxter}
\centering
\begin{tabular}{m{0.11\textwidth} m{0.11\textwidth} m{0.12\textwidth} m{0.07\textwidth} m{0.07\textwidth} m{0.17\textwidth} m{0.17\textwidth} m{0.12\textwidth} }
\midrule
\hline
\hline
\centering {\bf ID} & \begin{center}{\bf Sample} \end{center}  & \begin{center} {\bf Cartoon} \end{center}  & \begin{center} {{\bf $f_1$ ($f_{OTS}$)}} \end{center} &\begin{center} {{\bf $f_2$ ($f_{glass}$)}} \end{center}  &  \multicolumn{2}{c}{\bf Equilibrium contact angle } & \begin{center} {\bf Cassie-Baxter \\ prediction} \end{center}   \\
\hline
\begin{center} GLASS \end{center} & \begin{center} homogeneous \end{center} & \vspace{1mm}  \parbox[c]{1em}{\includegraphics[scale=0.108]{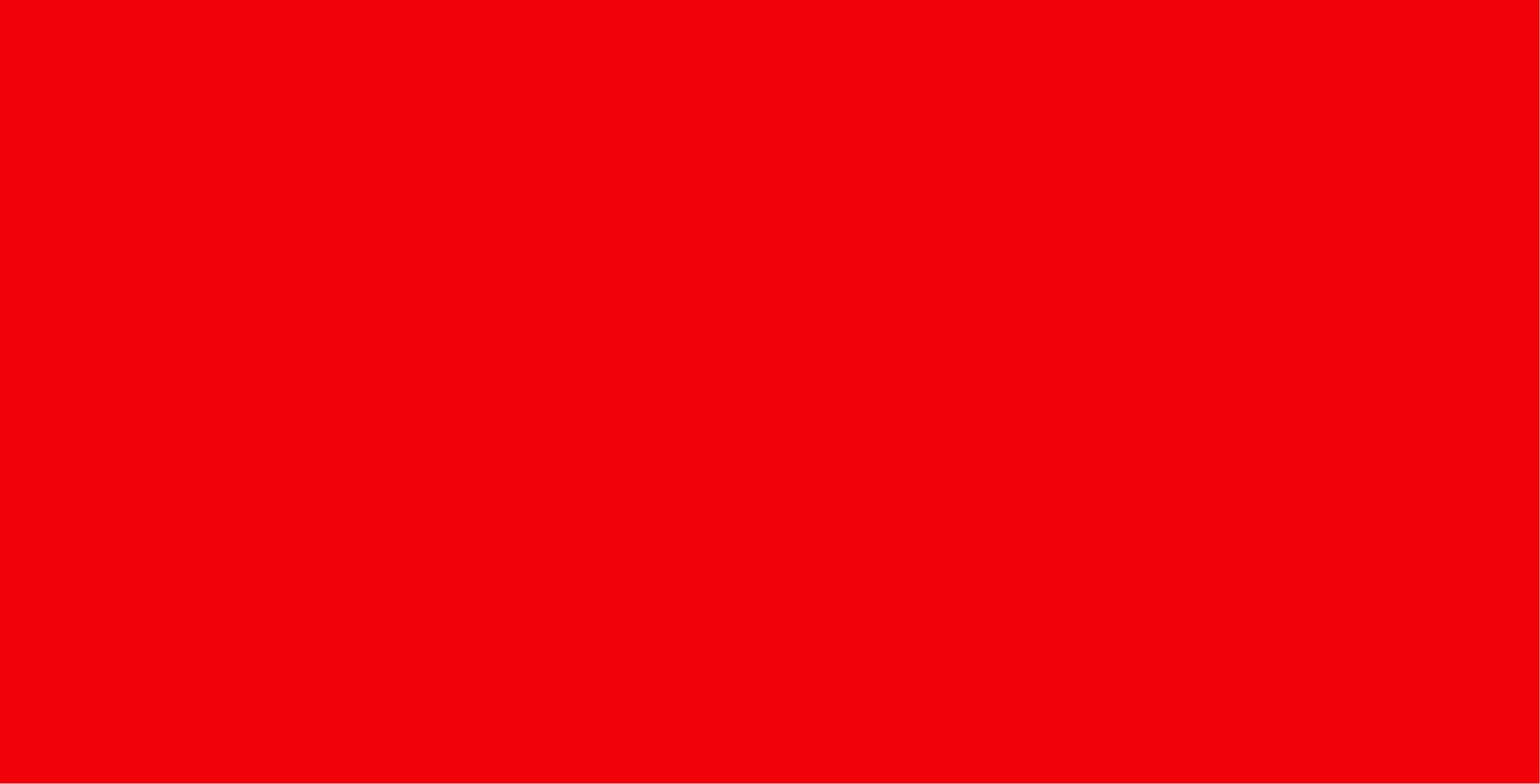} } &   \begin{center} 0  \end{center} &  \begin{center} 1 \end{center} &  \multicolumn{2}{c}{ $32^{\circ} \pm 3^{\circ}$} &  \begin{center} - \end{center}  \\
\begin{center} OTS \end{center} & \begin{center} homogeneous \end{center}  & \vspace{1mm} \parbox[c]{1em}{\includegraphics[scale=0.108]{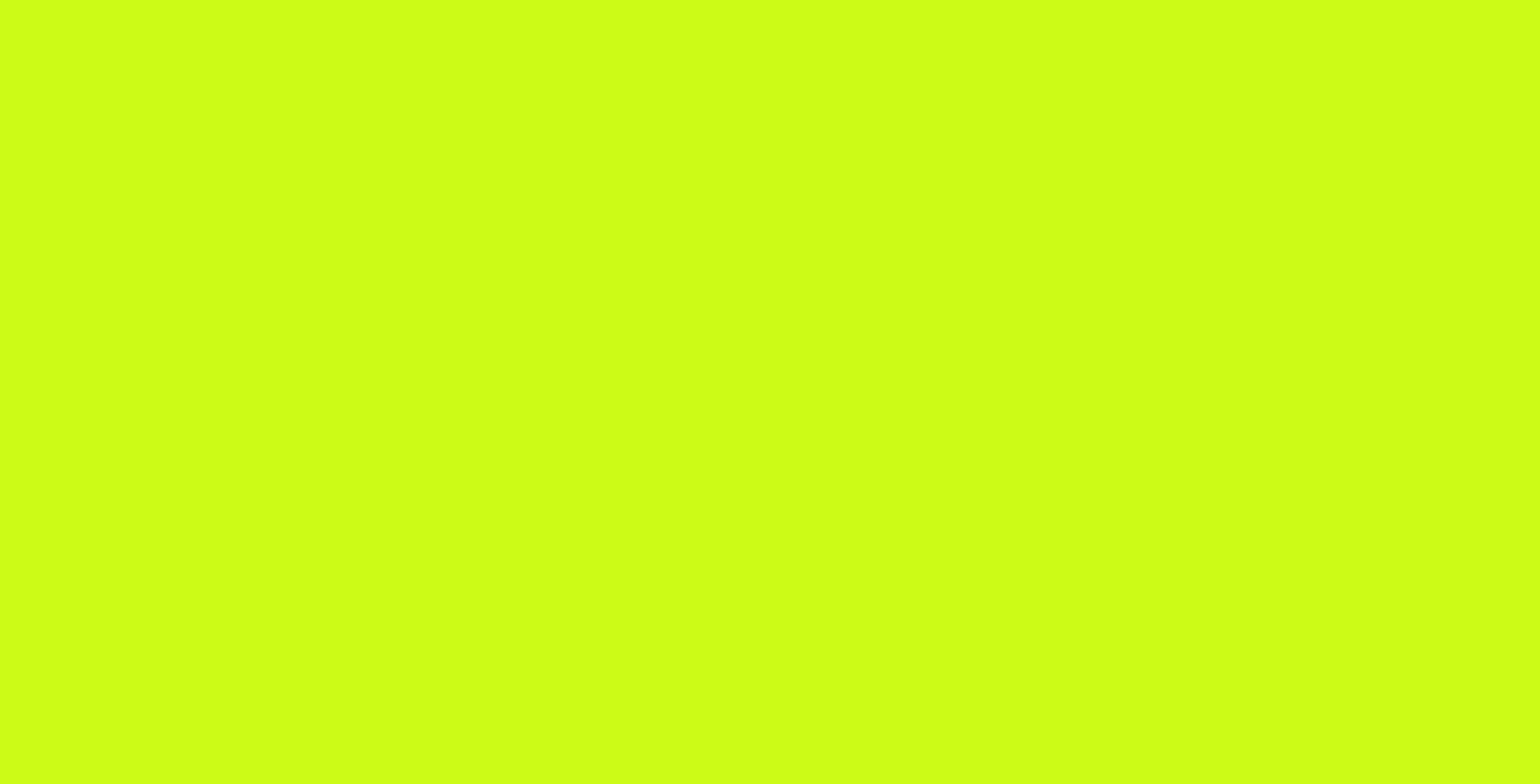}} \vspace{1mm}  & \begin{center} 1 \end{center}    & \begin{center} 0 \end{center} &   \multicolumn{2}{c}{$110^{\circ} \pm 3^{\circ}$} & \begin{center} - \end{center} \\  \hline
 &   &   &    &  &   \vspace{1mm} \parbox{1em}{\includegraphics[scale=0.32]{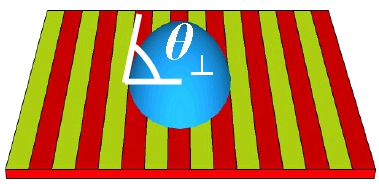}} \vspace{1mm} & \vspace{1mm} \parbox{1em}{\includegraphics[scale=0.31]{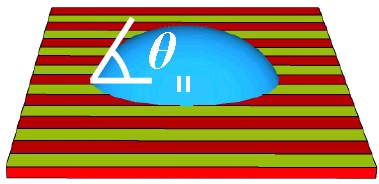}} \vspace{1mm} &  \\  \hline
\begin{center} \textcolor{black}{OTS\_19\%} \end{center} & \begin{center} heterogeneous \end{center} & \vspace{1mm} \parbox[c]{1em}{ \includegraphics[scale=0.108]{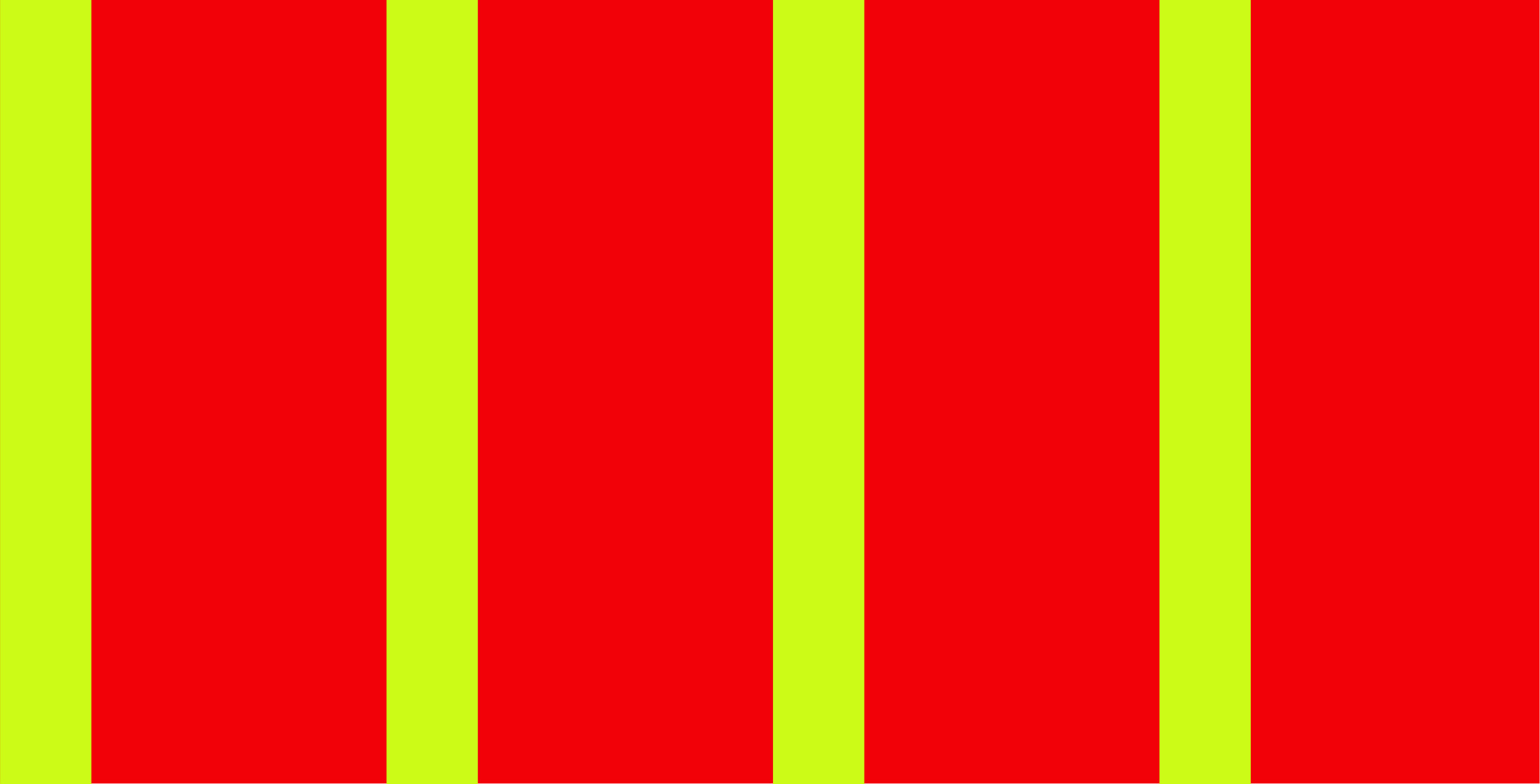}}     & \begin{center} 0.19 \\ \end{center} & \begin{center} 0.81 \end{center}  & \begin{center} $\theta _{\perp}=74^{\circ} \pm 3^{\circ}$ \end{center} & \begin{center} $\theta _{||}=58^{\circ} \pm 4^{\circ}$ \end{center} & \begin{center} $52^{\circ} \pm 2^{\circ}$ \end{center}  \\
\begin{center} \textcolor{black}{OTS\_50\%} \end{center} & \begin{center} heterogeneous \end{center} & \vspace{1mm} \parbox[c]{1em}{\includegraphics[scale=0.108]{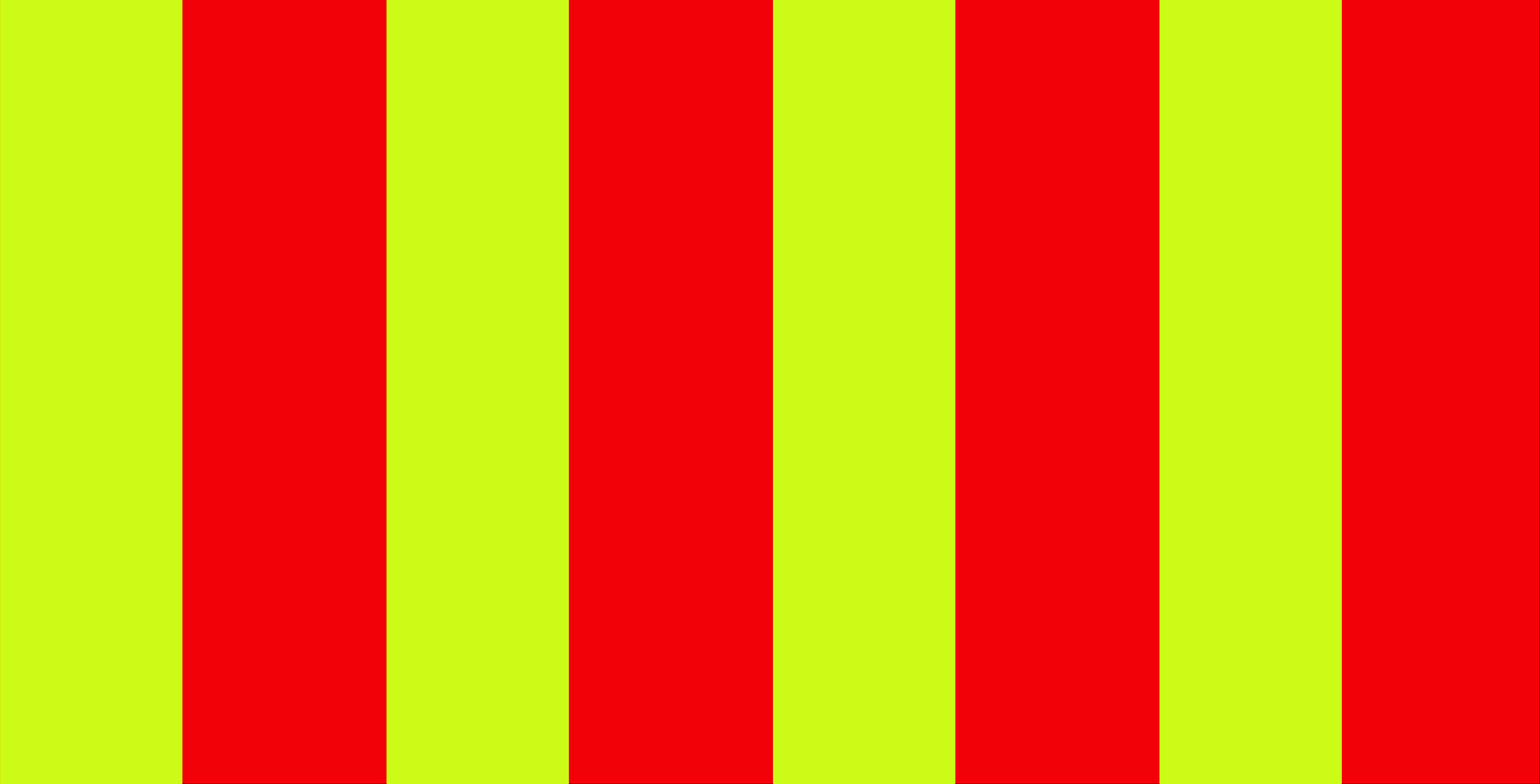} }    & \begin{center} 0.50 \end{center} & \begin{center} 0.50  \end{center}  & \begin{center} $\theta _{\perp}=83^{\circ} \pm 2^{\circ}$ \end{center} & \begin{center} $\theta _{||}=72^{\circ} \pm2^{\circ}$ \end{center} & \begin{center} $75^{\circ} \pm 2^{\circ}$ \end{center}  \\
\begin{center} \textcolor{black}{OTS\_83\%} \end{center} & \begin{center} heterogeneous \end{center} & \vspace{1mm} \parbox[c]{1em}{\includegraphics[scale=0.108]{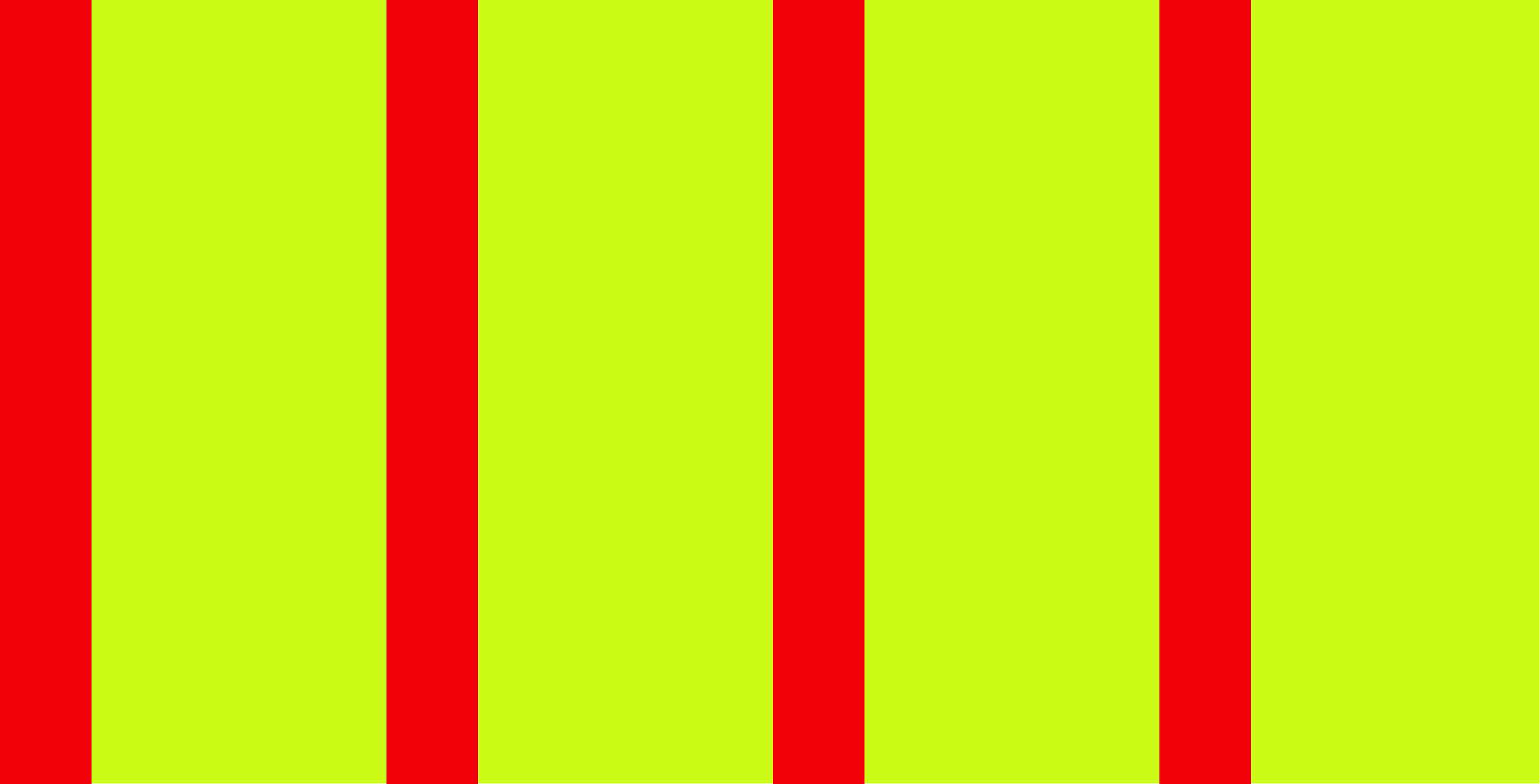}} \vspace{1mm}  & \begin{center} 0.83\end{center} & \begin{center} 0.17 \end{center}   & \begin{center} $\theta _{\perp}=100^{\circ} \pm 3^{\circ}$\end{center} & \begin{center} $\theta _{||}=100^{\circ} \pm 3^{\circ}$ \end{center} & \begin{center} $98^{\circ} \pm 2^{\circ}$ \end{center}   \\
\hline
\hline
\end{tabular}
\end{table*}


\subsection{Experimental Results}\label{sec:1b}

\begin{figure}[h!]
\includegraphics[scale=1]{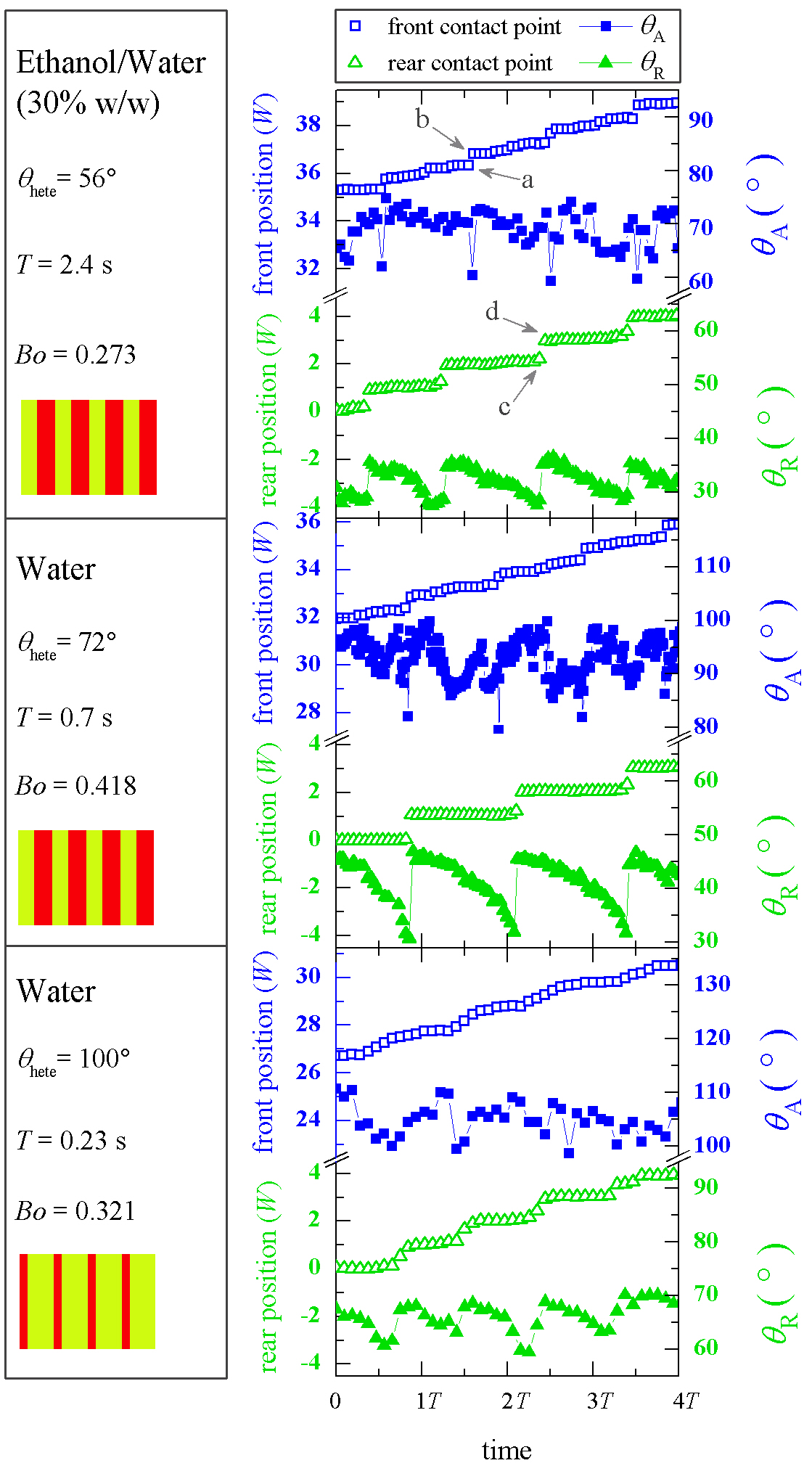}
\caption{(Color online) Time dependence of the front and rear contact points (left axis) and the advancing and receding contact angles (right axis). Space is expressed in units of the pattern periodicity $W$ and time in units of the period $T$ (the time required for a displacement of the drop equal to $W$). Top panel: measurement about a 30 $\mu\textrm{l}$ drop of ethanol in water (30\% w/w) sliding down the sample with stripes of equal width (\textcolor{black}{OTS\_50\%}); middle panel: data about a 30 $\mu\textrm{l}$ water drop on the surface \textcolor{black}{OTS\_50\%} formed by the same percentage of glass and OTS ($f_1=f_2=0.5$); bottom panel: measurement about a 30 $\mu\textrm{l}$ water drop on the sample \textcolor{black}{OTS\_83\%} featuring larger stripes of OTS (the hydrophobic part) with $f_1=0.83$, $f_2=0.17$. Boxes on the left report corresponding experimental details.\label{stick-slip}}
\end{figure}

The sliding of water drops down the heterogeneous samples has been observed in the direction perpendicular to the stripes, as shown in Fig.~\ref{pianoinclinato}. In the case of surfaces with wider stripes of glass (Fig.~\ref{condensation}a), drops assume an asymmetric shape, elongated in the direction of the stripes, and get pinned for every inclination angle up to $90^{\circ}$ so that sliding measurements are not possible. Drops on surfaces with stripes of glass and OTS of equal width (Fig.~\ref{condensation}b) and on surfaces with larger stripes of OTS (\textcolor{black}{Fig.}~\ref{condensation}c) are not affected by this pronounced asymmetry and the motion is studied for various inclinations $\alpha$ of the sample. To extend the range of static wettability on the heterogeneous samples, we also studied the sliding of ethanol in water drops (see Sec.~\ref{sec:1a}) down the surface with stripes of equal width. An example of the particular drop dynamics in these three different situations is shown in Fig.~\ref{stick-slip}. The drop clearly advances with a stick-slip behavior, with jumps of the order of the pattern periodicity $W$, on the surface formed by OTS and glass stripes of equal width (see the upper and middle panels of Fig.~\ref{stick-slip}). The time period $T$ is defined as the time required to a drop for a displacement equal to $W$. Considering point (a) in the top graph of Fig.~\ref{stick-slip} as the beginning of $T$, at point (b) the front of the drop suddenly jumps forward by a distance almost equal to $W/2$, while the rear contact line is pinned.  After the jump, the front line slowly advances and subsequently the rear line jumps by a distance equal to $W$, corresponding to points (c) and (d). The period $T$ ends when the front contact point covers a length of $W/2$ before performing the next jump. The process then repeats itself. In correspondence to the leap of the front line, a fall in $\theta_{A}$ occurs, whereas $\theta_{R}$ reaches the minimum value just before the depinning of the rear contact point, then jumps to the maximum value in correspondence of the crossing of $W$ and finally, during the subsequent pinning, gradually decreases. We point out that the pinning-depinning transition occurs through a discontinuity both in the position and in the contact angle resulting more pronounced in the case of the rear of the drop. This behavior is observed both in the case of ethanol in water and pure water drops on the same surface (\textcolor{black}{OTS\_50\%}), differing only by the contact angle values that are higher in the case of water drops. On the other hand, the behavior of water drops on surfaces with larger stripes of OTS is quite different (see the bottom panel of Fig.~\ref{stick-slip}): even if drop motion is characterized by the same space periodicity $W$, the trend of the front and the rear contact points is smoother and does not feature any net jump. Also $\theta_{A}$ and $\theta_{R}$ exhibit only oscillations without any marked discontinuity.

\begin{figure}[!htbp]
\includegraphics[scale=1]{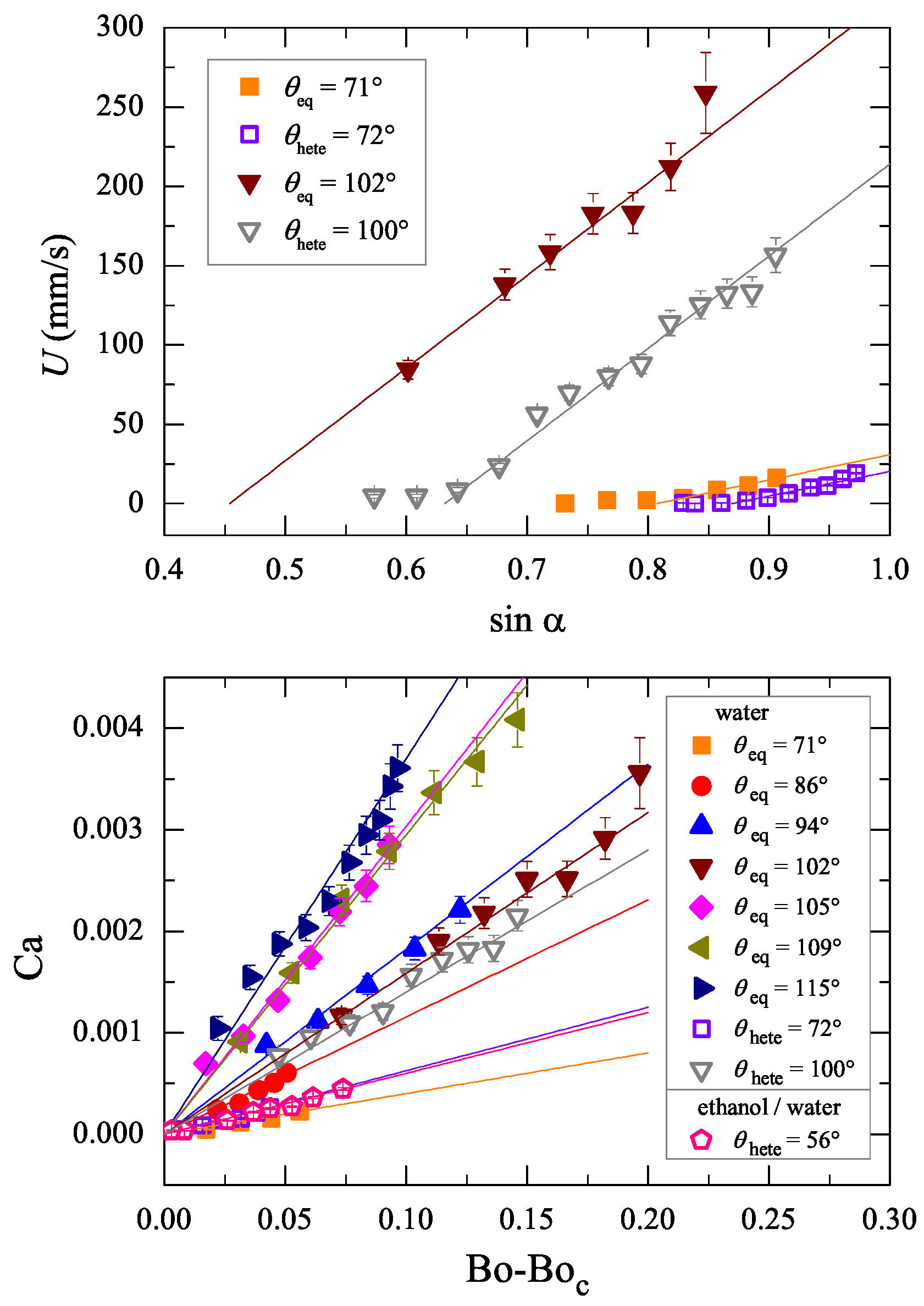}
\caption{(Color online) Top panel: mean velocity of 30 $\mu\textrm{l}$ water drops sliding down the heterogeneous (open symbols) surfaces \textcolor{black}{OTS\_50\%} ($f_1=f_2=0.5$, $\theta_{hete}=72^{\circ}$) and \textcolor{black}{OTS\_83\%} ($f_1=0.83$, $f_2=0.17$, $\theta_{hete}=100^{\circ}$)  and down homogeneous (filled symbols) surfaces of similar wettabilities, inclined by several angles $\alpha$. Lines are linear fit to the data taken on a range where viscous dissipation is not negligible. The intercept at $U=0$ defines the critical angle $\alpha_{c}$. Bottom panel: \textcolor{black}{Ca} {\it vs.} (\textcolor{black}{$\rm{Bo-Bo_c}$}) curves for homogeneous (filled symbols) and heterogeneous (open symbols) surfaces of different wettability. Measures on heterogeneous surfaces (open symbols) are performed with water drops on samples \textcolor{black}{OTS\_50\%} ($f_1=f_2=0.5$) and \textcolor{black}{OTS\_83\%} ($f_1=0.83$, $f_2=0.17$) and with ethanol in water (30\% w/w) drops on sample \textcolor{black}{OTS\_50\%} ($f_1=f_2=0.5$), as reported in Fig.~\ref{stick-slip}. The experimental data and the corresponding fits are horizontally shifted by \textcolor{black}{$\rm{Bo_{c}}$}.  \label{U_Ca_Bo}}
\end{figure}

By performing sliding measurements we can derive the relationship between the drop mean velocity $U$ and the inclination angle $\alpha$ of the surface. Fig.~\ref{U_Ca_Bo} reports data of water drops sliding on striped surfaces \textcolor{black}{OTS\_50\%} and \textcolor{black}{OTS\_83\%} and on homogeneous surfaces with similar wettabilities. Above the critical angle $\alpha_{c}$ the sliding velocity $U$ scales linearly with $\sin \alpha$, as described by Eq. (\ref{eq:scaling}). We point out that experimentally we still observe motion even for tilt a few degrees ($\lesssim 5^{\circ}$) smaller than $\alpha_{c}$, a condition in which the drop is moving at low \textcolor{black}{Ca} where the viscous dissipation is negligible and the prediction of Eq.~(\ref{eq:scaling}) is no more applicable. Nonetheless the determination of $\alpha_c$ has been performed by extrapolating the linear trend in the dissipative sliding up to zero velocity. Indeed the stick-slip regime is typically well observed close to $\alpha_c$. Considering the heterogeneous and homogeneous surfaces with similar equilibrium contact angle, we observe two distinctive features:  $i)$  at the same inclination $\alpha$, the velocity is always lower on the heterogeneous surface than on the homogeneous one and the angle $\alpha_{c}$ is higher for the heterogeneous surfaces which are characterized by a larger pinning; $ii)$ the slope of the curve $U$ {\it vs.} $\sin \alpha$ is the same for similar wettability, regardless of the composition of the surface, and is higher for the surfaces characterized by higher equilibrium contact angle. To better understand the dependence of the curve $U$ {\it vs.} $\sin \alpha$ on the static wettability, we extended these measurements to several homogeneous samples featuring different equilibrium contact angles. Such data are collected in the bottom panel of Fig.~\ref{U_Ca_Bo} and expressed in terms of the dimensionless numbers \textcolor{black}{Ca} and \textcolor{black}{$\rm{Bo-Bo_{c}}$} in order to better appreciate the range of slopes of the curves. We underline how the slope \textcolor{black}{$\Delta \rm{Ca}/ \Delta \rm{Bo}$}, being inversely proportional to the dissipation (see Sec.~\ref{sec:1}), clearly increases as the hydrophobicity of the surfaces increases~\cite{HuHScriven71,Kimetal02}. 

\section{Numerical Results}\label{sec:2}

\begin{figure}[t]
\includegraphics[scale=0.4]{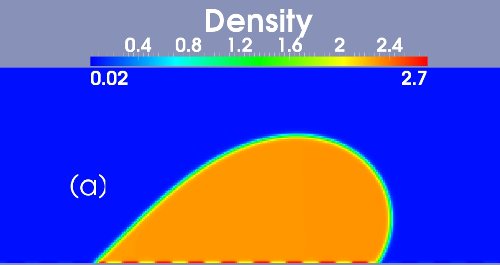}
\includegraphics[scale=0.4]{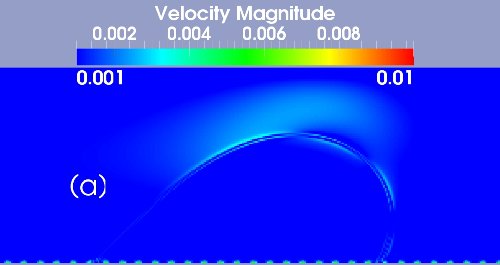}
\includegraphics[scale=0.47]{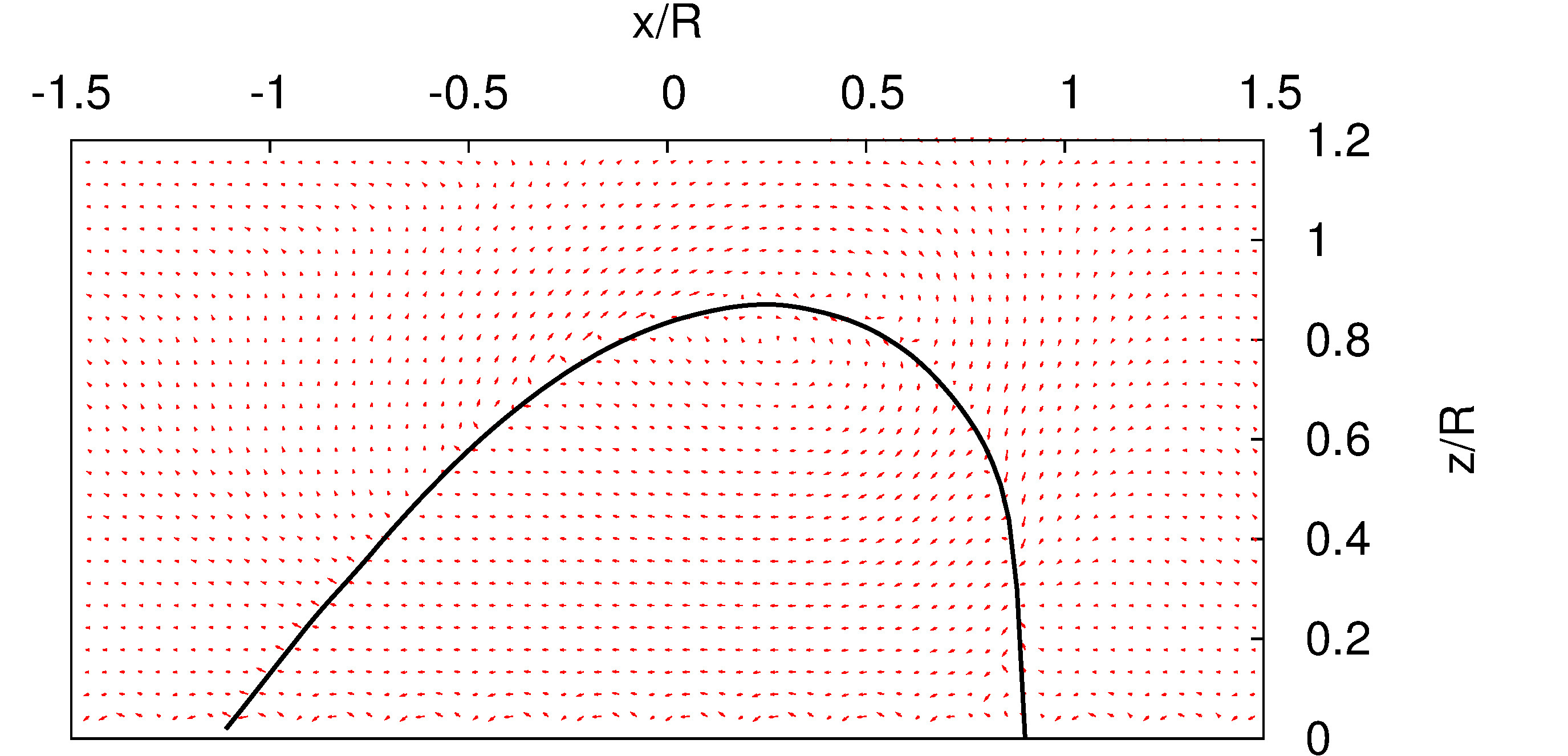}\\
\includegraphics[scale=0.4]{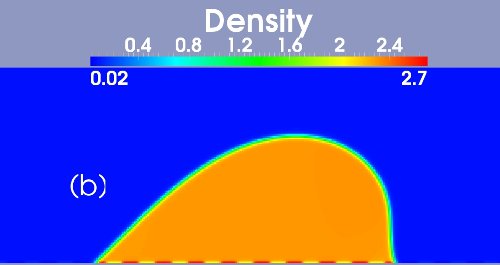}
\includegraphics[scale=0.4]{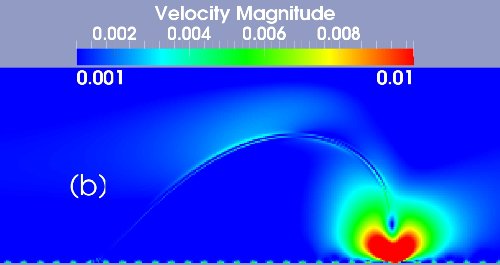}
\includegraphics[scale=0.47]{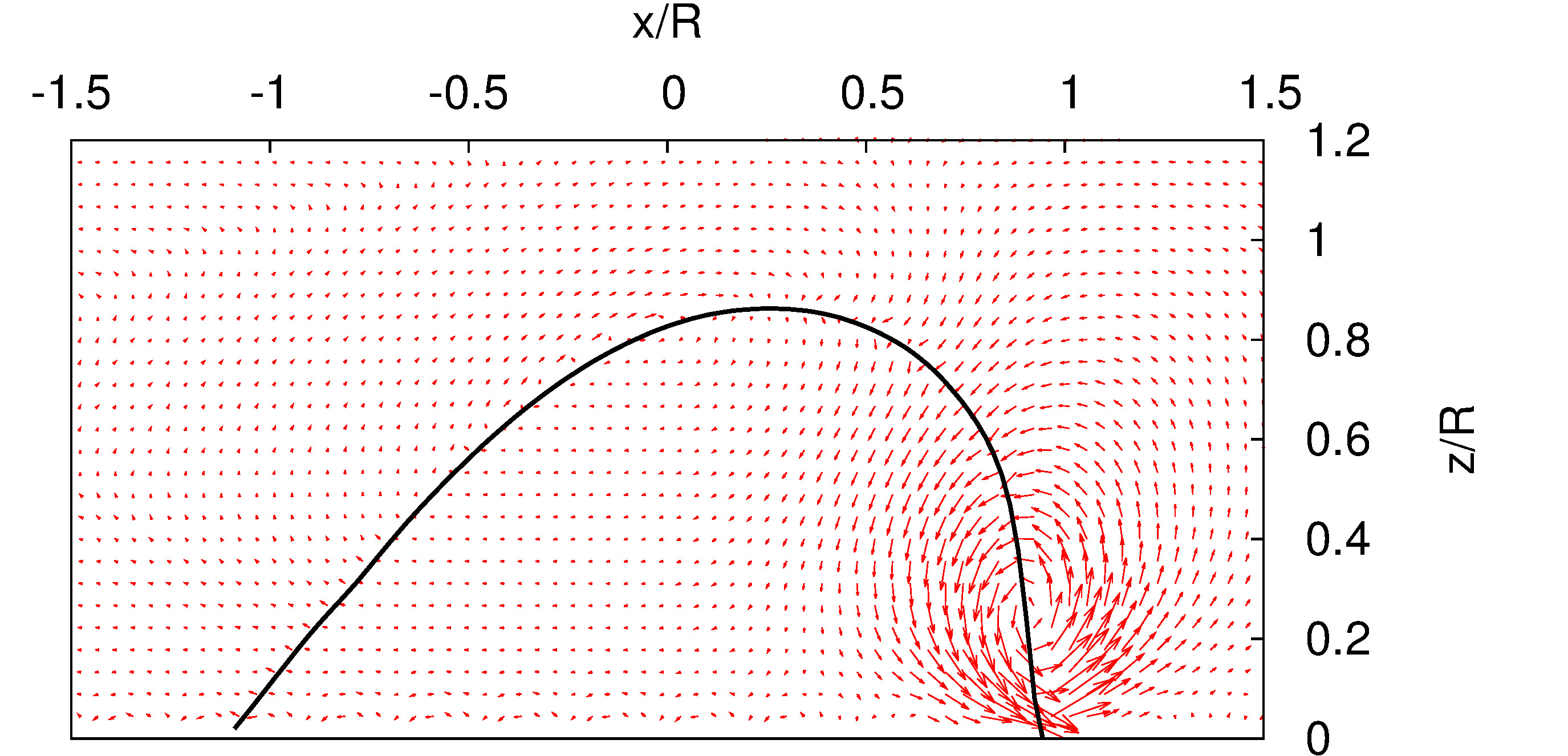}\\
\includegraphics[scale=0.4]{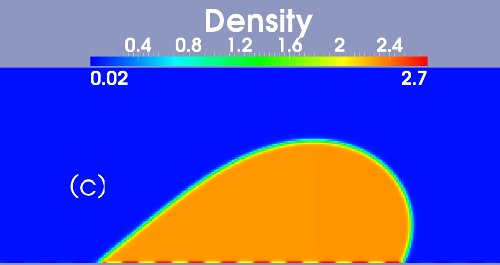}
\includegraphics[scale=0.4]{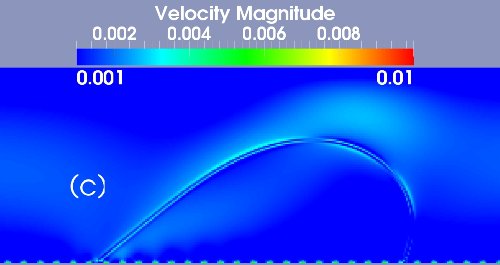}
\includegraphics[scale=0.47]{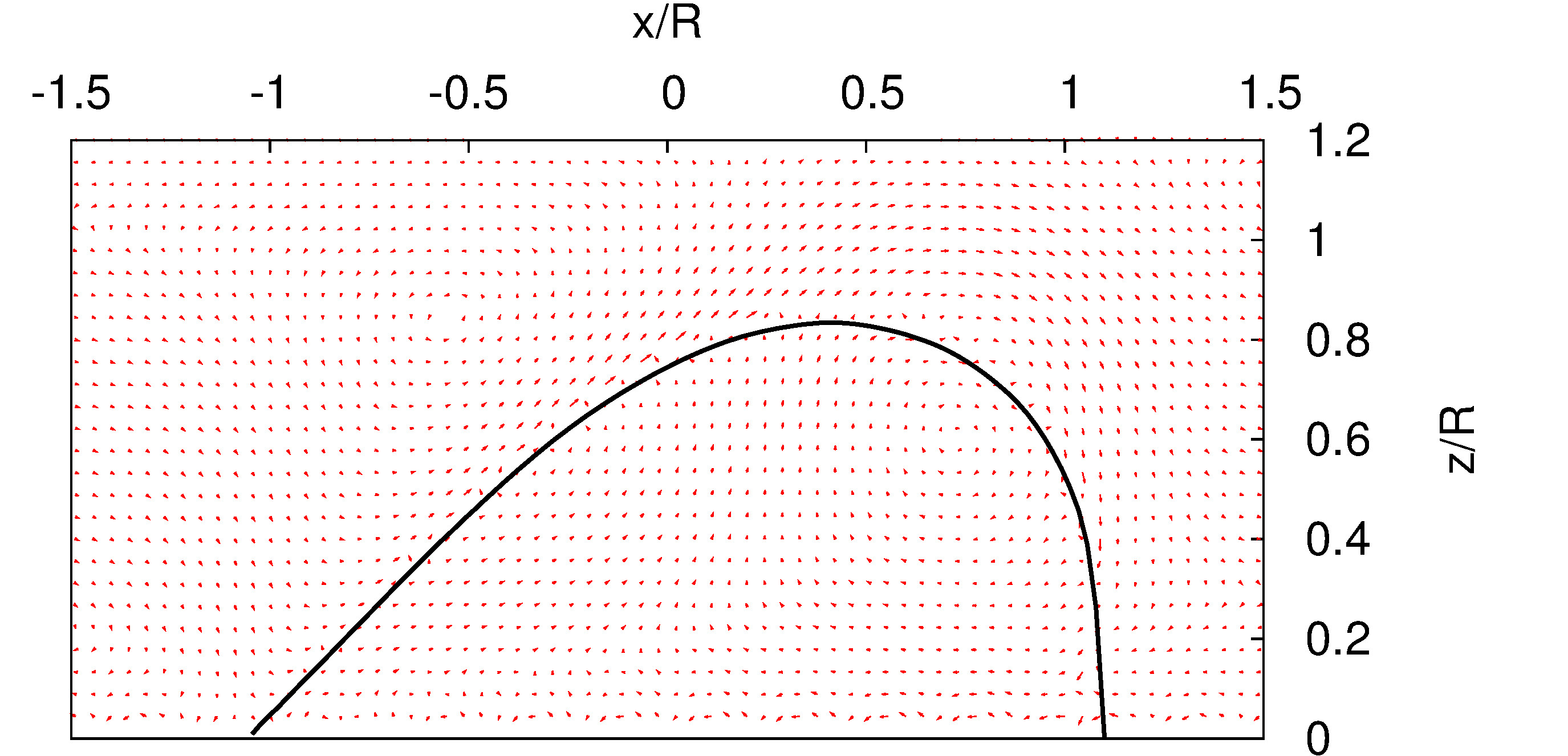}\\
\includegraphics[scale=0.4]{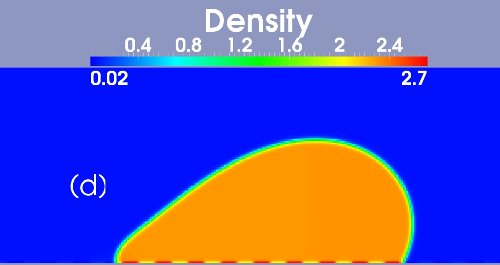}
\includegraphics[scale=0.4]{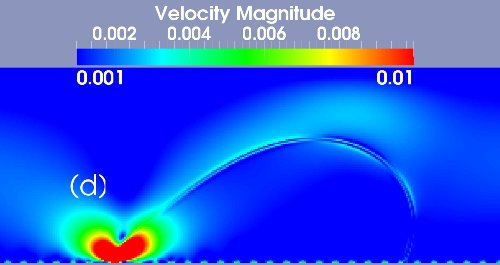}
\includegraphics[scale=0.47]{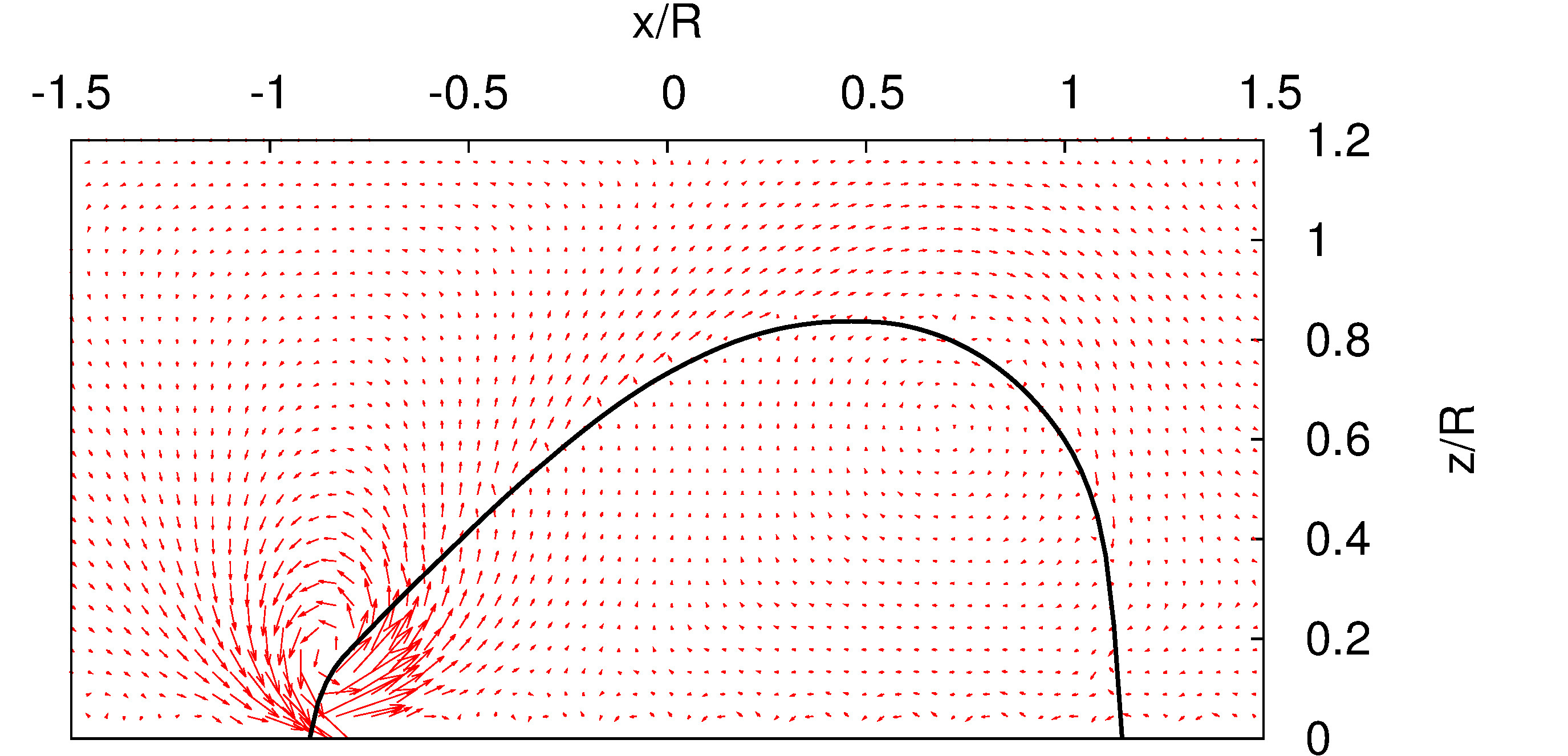}\\
\caption{\textcolor{black}{(Color online) Density snapshots (left column) during the stick-slip dynamics for a situation with the same fraction of hydrophilic and hydrophobic areas, i.e. $f_1=f_2=0.5$, and for a Bond number Bo=0.017. The orange/light (blue/dark) color is associated to high (low) density regions. The four snapshots (a)-(d) refer to four different time steps as reported in the top panel of Fig.~\ref{fig:SLnum}. The corresponding velocity-magnitude snapshots (middle column) are also reported. All data are reported in lbu (LB units). In the right column panel we report Momentum field inside drop in the center-of-mass frame.\label{fig:densvel}}}
\end{figure}

\begin{figure}[tbp]
\includegraphics[scale=0.6]{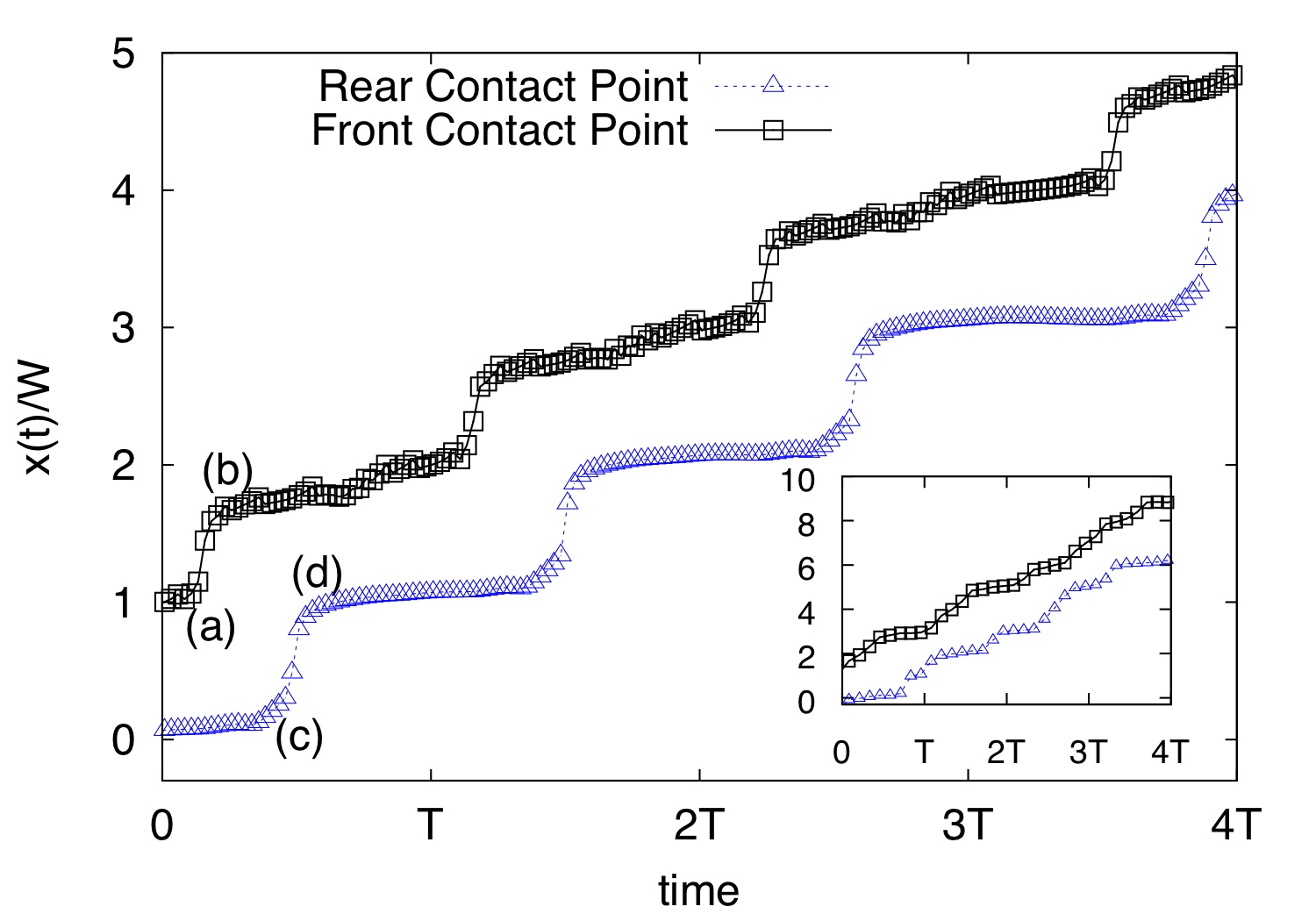}\\
\includegraphics[scale=0.61]{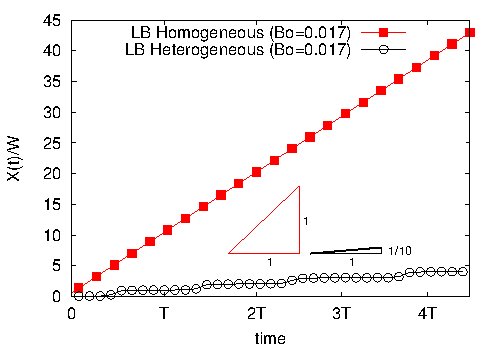}
\caption{(Color online) Top panel: time evolution in dimensionless units (see text for details) of the position of the rear and the front contact points of a drop sliding down the heterogeneous surface for a Bond number \textcolor{black}{Bo=0.017} and $f_1=f_2=0.5$. The front contact point position is translated for visualization. Letters refer to the snapshots for density and velocity magnitude reported in Fig.~\ref{fig:densvel}. The inset shows the corresponding situation with larger hydrophobic stripes, achieved by simulating a case with $f_1=0.75$, $f_2=0.25$. The bottom panel compares the position of the rear contact point for the heterogeneous ($f_1=f_2=0.5$) and the homogeneous case at the same Bond number (\textcolor{black}{Bo=0.017}), with the homogeneous equilibrium contact angle chosen in agreement with the Cassie-Baxter equation (\ref{eq:CB}). The time scale $T$ indicates the characteristic period of the stick-slip dynamics at \textcolor{black}{Bo=0.017}. The average speed of the heterogeneous case is visibly an order of magnitude less than that of the homogeneous case. \label{fig:SLnum}}
\end{figure}

For the numerical simulations we employ a mesoscopic LB model~\cite{Benzi92} to reproduce the diffuse interface dynamics of a binary mixture. LB turned out to be a very effective method to  describe mesoscopic physical interactions and non-ideal interfaces coupled to hydrodynamics. Many multiphase and multicomponent LB models have been developed, on the basis of different points of view, including the Gunstensen model~\cite{Gunstensen}, the free-energy model~\cite{YEO} and the ``Shan-Chen'' model~\cite{SC}, the latter being widely used thanks to  its simplicity and efficiency in representing interactions between different species and different phases~\cite{Kupershtokh,CHEM09,Sbragaglia07,Shan06b,Sbragagliaetal12,VarnikSaga,JansenHarting11}. The numerical simulations with the LB models (see appendix) are used to reveal the importance of the various terms in the equations of motion. In particular, these numerical simulations are crucial to elucidate the relative importance of capillary, viscous and body forces in the dynamical evolution of the drop. We will analyze the case of a cylindrical drop on a chemically striped surface with the drop radius such that $R \approx 10 W$. Simulating two-dimensional drops allows to better resolve the hydrodynamics inside the drop and approach with higher accuracy the hydrodynamic limit of the LB equations (see Appendix). We will first present results with a  viscous ratio $\chi=\eta_{in}/\eta_{out}=1$, where $\eta_{in}$, $\eta_{out}$ are the dynamic viscosities inside (inner viscosity) and outside (outer viscosity) the drop, respectively. Later on, we will also specialize to the case of different dynamic viscosities, to better compare with the experimental results. The dynamic equations we reproduce are the continuity equations and the Navier-Stokes equations of a fluid mixture with two components $\zeta=A,B$, with $A$ the rich component in the drop phase. As for the momentum equation, in the limit of very small Reynolds number, we integrate in time the following equation ($x$ is the down plane coordinate and repeated indexes are summed upon)
\be\label{NS}
\rho \frac{\partial u_{x}}{\partial t} = -\frac{\partial P_{x \beta}}{\partial r_{\beta}}+ \frac{\partial \sigma_{x \beta}}{\partial r_{\beta}} + \rho_A g \sin \alpha \delta_{ix}
\ee
where $\rho_{\zeta}$ is the density of the $\zeta$-th component ($\rho=\sum_{\zeta} \rho_{\zeta}$ is the total density), ${u}_{\alpha}$ refers to the $\alpha$-th projection of the fluid velocity, $\sigma_{\alpha \beta}$ is the viscous stress tensor and  $P_{\alpha \beta}$ is the pressure tensor~\cite{SbragagliaBelardinelli} encoding both the non-ideal effects at the interface (liquid-gas surface tension) and the interation with the solid wall (wettability).  All the details of the model are reported in the Appendix. The diffuse interface time-dependent Stokes equation (\ref{NS}) is integrated over the drop volume and made dimensionless with respect to the surface tension force $R\gamma_{LG}$.  We end up with the following balance
\begin{equation}
M a(t)=F_{cap}(t)+D(t)+F_{g}  \label{eq:balance}
\end{equation}
where $a(t)$ is the acceleration of the drop with mass $M$ and $F_{g}$ is the down-plane component of the gravitational force. The term $F_{cap}$ (calculated as the integral of the pressure tensor term) accounts for the nonuniform pressure and curvature distortion as well as the capillary force on the drop at the contact line. The function $D(t)$ (the integral of the viscous stress term) quantifies the drag force due to viscous shear.\\
In Figs.~\ref{fig:densvel} and~\ref{fig:SLnum} we show the emergence of the stick-slip dynamics in the numerical simulations. We have reproduced the same wettabilities experimentally investigated in Fig.~\ref{stick-slip} and explored different values of the Bond numbers by changing the value of $g \sin \alpha$. 
Figure~\ref{fig:densvel} reports snapshots of the density and velocity  corresponding to the pinning and depinning transition of the drop. In the left sequence (density snapshots), the front contact line gets pinned before entering the hydrophobic regions (snapshot (a)). Then it penetrates slowly through the hydrophobic area with an increasing advancing angle until it enters the hydrophilic region performing a sudden jump (snapshot (b)). The rear contact line motion on the hydrophilic and hydrophobic stripes is similar, the only difference being the receding contact angle is reducing as the drop stays pinned, and increases after the jump (snapshots (c) and (d)). \textcolor{black}{In parallel, the middle \textcolor{black}{sequence} of velocity snapshots shows a velocity magnitude close to zero during the pinning on hydrophobic areas (snapshots ($\textrm{a}$) and ($\textrm{c}$)) and a spike in the correspondence of the drop slip (snapshots ($\textrm{b}$) and ($\textrm{d}$)). In the right \textcolor{black}{sequence} we report the momentum field in (and around) the drop in the reference frame of the center of mass. In a stationary homogeneous case \textcolor{black}{(not shown)}, we confirm  the presence of a well established rotational flow \cite{Moradi,Thampi,Servantie,Mognetti}.}  \textcolor{black}{On the other hand, the sliding on heterogeneous surfaces is characterized by rotational flow mostly near the depinning contact point, as we can see from the snapshots corresponding to the rear and front jumps (see Fig.~\ref{fig:densvel}).} We point out that the jumps of the front and rear contact lines do not take place at the same instant, since the front sticks as the rear slips and {\it vice versa}, as clearly confirmed both experimentally (Fig.~\ref{stick-slip}) and numerically (Fig.~\ref{fig:SLnum}). Correspondingly, the top panel of Fig.~\ref{fig:SLnum}  displays the time evolution of the positions of the front and rear contact points normalized to $W$, for a situation with the same fraction of hydrophilic and hydrophobic areas, i.e. $f_1=f_2=0.5$, and for a Bond number \textcolor{black}{Bo}=0.017~\cite{prl13}. The time lag $T$ is the characteristic period of the stick-slip dynamics, similarly to what is reported in Fig.~\ref{stick-slip}. In the inset of the top panel of Fig.~\ref{fig:SLnum} we can appreciate the change in the dynamics induced by larger hydrophobic stripes,  achieved by simulating a case with $f_1=0.75$, $f_2=0.25$: drop motion has the same space periodicity $W$, but the front contact point motion is smoother, similarly to what we have experimentally observed in the bottom panel of Fig.~\ref{stick-slip}. The rear contact point, instead, experiences more frequent jumps forward. This may be seen as a signature of the transition from the regular stick-slip dynamics to a homogeneous stationary motion. In the bottom panel of Fig.~\ref{fig:SLnum} we compare the stick-slip dynamics of the heterogeneous case with $f_1=f_2=0.5$ with that of a homogeneous substrate at the same Bond number (\textcolor{black}{Bo=0.017}), with the homogeneous equilibrium contact angle chosen in agreement with the Cassie-Baxter equation (\ref{eq:CB}). The mean velocity of the heterogeneous case is visibly an order of magnitude less that that of the homogeneous case.

\begin{figure}[tbp]
\includegraphics[scale=0.5]{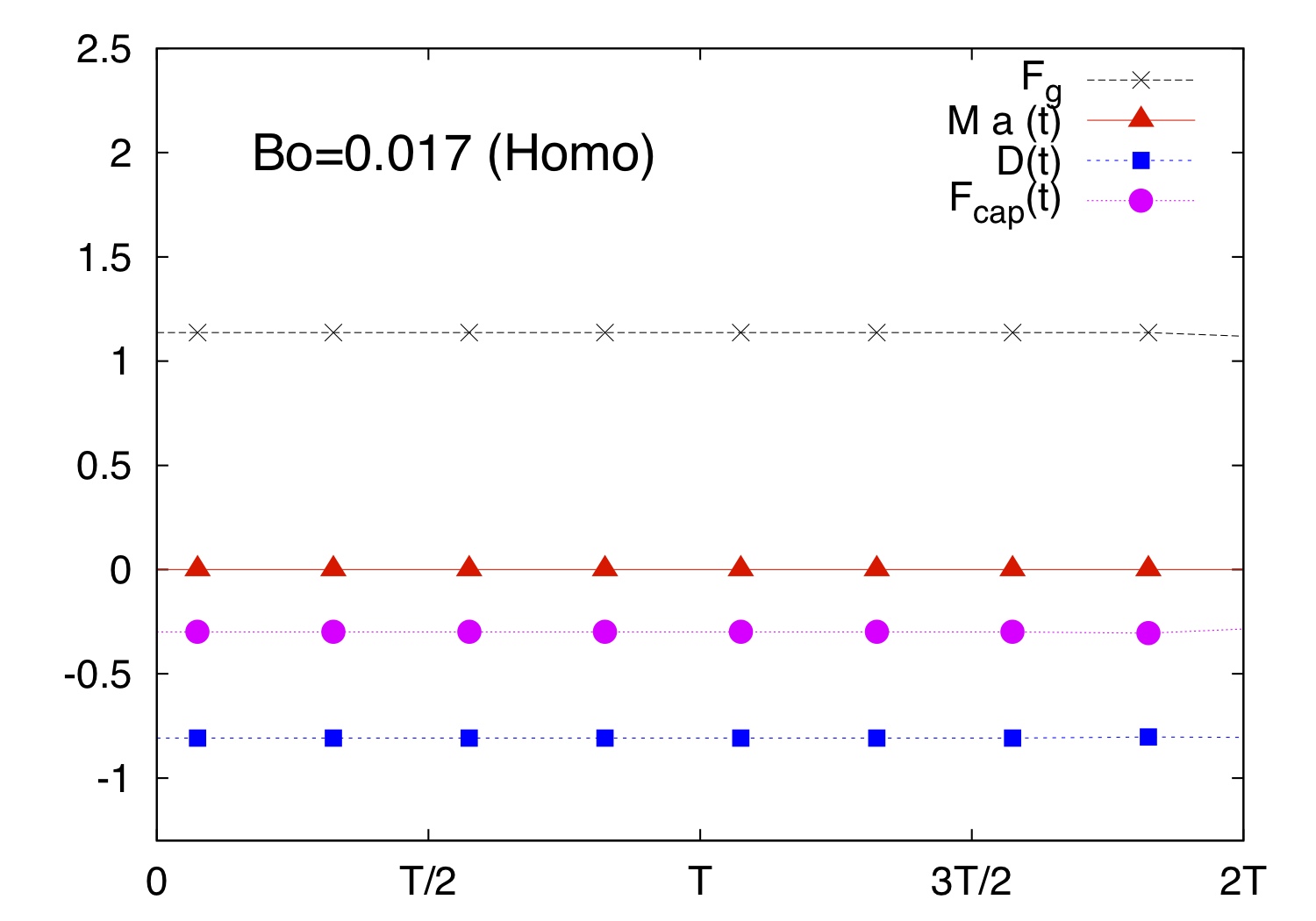}\\
\includegraphics[scale=0.5]{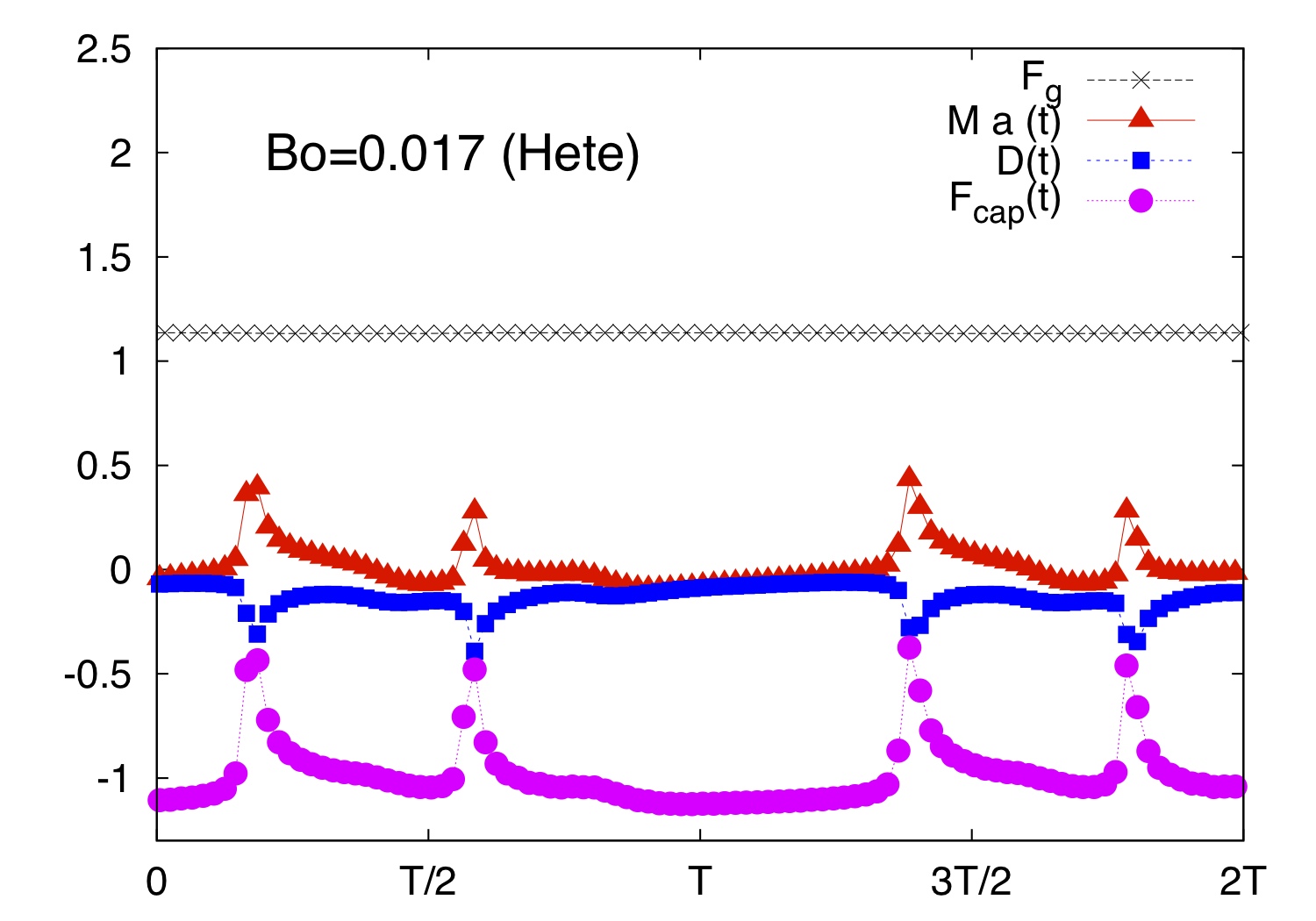}\\
\includegraphics[scale=0.5]{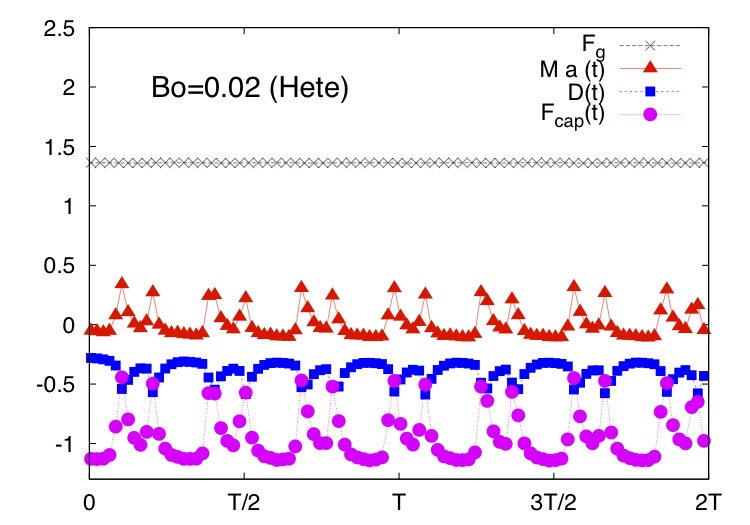}
\caption{(Color online) Time evolution of the various terms in the balance equation (\ref{eq:balance}) for the homogeneous and heterogeneous ($f_1=f_2=0.5$) cases. Both the top panel (homogeneous case with static angle $\theta_{eq} \approx 85^{\circ}$) and the middle panel (heterogeneous case with  $\theta_{hete} \approx 85^{\circ}$, with $\theta_1=50^{\circ}$, $\theta_2=120^{\circ}$)  refer to \textcolor{black}{Bo=0.017}. The bottom panel shows the effect of an increase of the Bond number on the dynamics for the heterogeneous pattern at \textcolor{black}{$\rm{Bo=0.02}$}. The time lapse considered is the same for the three cases, with the time scale $T$ indicating the characteristic period of the stick-slip dynamics at \textcolor{black}{$\rm{Bo=0.017}$}.}
\label{fig:EQMOTION}
\end{figure}

Fig.~\ref{fig:EQMOTION} presents the analysis of the balance equation (\ref{eq:balance}), comparing the sliding on homogeneous and heterogeneous surfaces, for the same \textcolor{black}{Bo} and for a time frame $2T$. The homogeneous case (top panel) is steady: the energy provided by $F_g$ is almost entirely transferred into dissipation, apart from the deformation of the interface which causes a term $F_{cap}$ smaller by a factor $\approx $10 with respect to the heterogeneous case (middle panel). As already reported~\cite{prl13}, in the striped surface, when the drop is pinned, $F_g$ is almost balanced by $F_{cap}$ (time step (a) in Fig.~\ref{fig:densvel}). Immediately after, the front contact line jumps forward  and the drop depins ($F_{cap} \rightarrow 0$) with a consistent dip in the viscous drag force (time step (b) in Fig.~\ref{fig:densvel}). The process repeats itself for the rear contact line  (time steps (c) and (d) in Fig.~\ref{fig:densvel}). Overall, we see that the effective dissipation in the heterogeneous case is strongly suppressed as compared with the stationary homogeneous case.  This is because the large wettability contrast causes additional energy to be stored in the non-equilibrium configuration of the drop which can pin before the contact lines jump forward. The analysis of the balance equation (\ref{eq:balance}) helps also to understand the transition from the stick-slip dynamics to the steady motion. The bottom panel of Fig.~\ref{fig:EQMOTION} shows the effect of an increase in the Bond number for the dynamics on the heterogeneous case, with the time scale $T$ still indicating the characteristic period of the stick-slip dynamics at \textcolor{black}{Bo=0.017}: as the Bond number is increased, the jumps of the rear and the front contact points become more frequent while the amplitude of the fluctuations of $F_{cap}$ and the acceleration $a(t)$ do not change appreciably. The change in \textcolor{black}{Bo} is compensated by an increase of the drag force, and hence, an increase of the mean velocity. For even larger \textcolor{black}{Bo} the drag force will dominate over $F_{cap}$, the variations in $a(t)$ and $F_{cap}$ become negligible, and the motion of the drop can be paralleled to that of a drop over a homogeneous substrate with an {\it effective} equilibrium contact angle (see below). \textcolor{black}{We point out that, when the Bo is significantly greater than $\rm{Bo_c}$ the relative contribution of the terms in Eq.~\ref{eq:balance} becomes more similar to the homogeneous case.}

\begin{figure}[!tbp]
\includegraphics[scale=0.6]{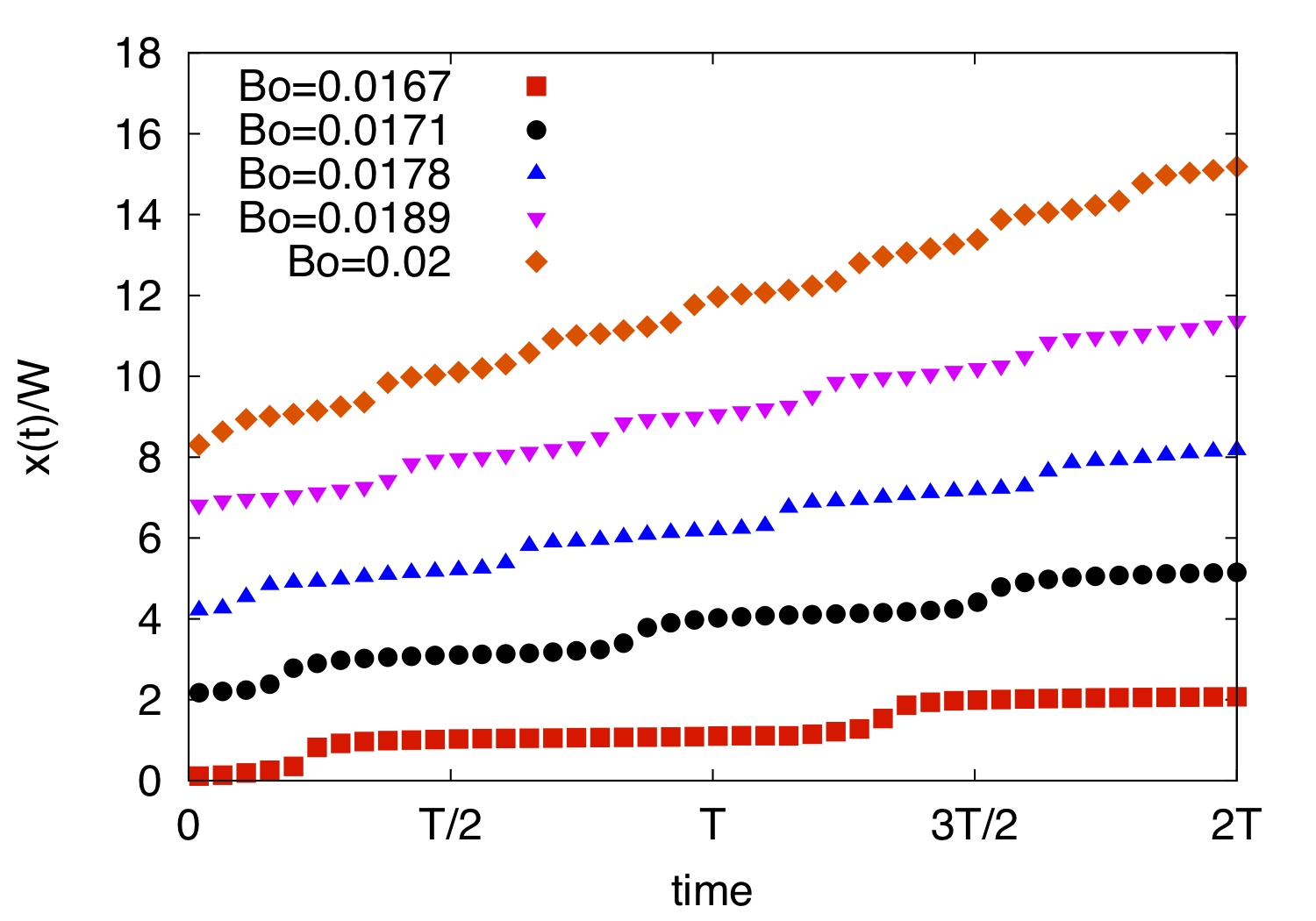}\\
\hspace{-.2in}\includegraphics[scale=0.65]{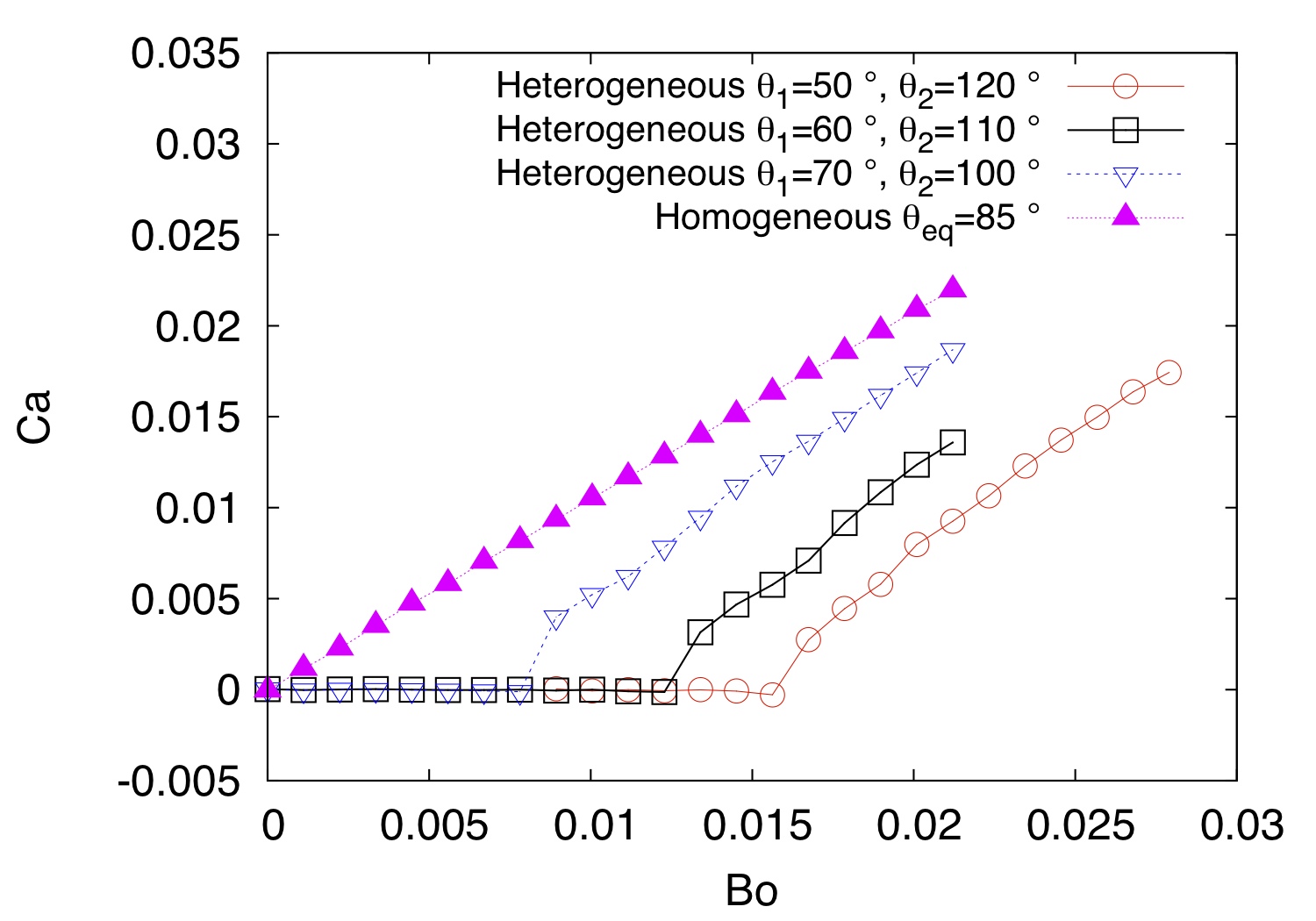}
\caption{(Color online) Top panel: time evolution in dimensionless units (see text for details) of the position of the front contact points for different \textcolor{black}{$\rm{Bo}$}. Bottom panel: relation between \textcolor{black}{Ca} and \textcolor{black}{$\rm{Bo}$} for both homogeneous and heterogeneous surfaces. In the heterogeneous case the capillary number is computed from the mean velocity and different intrinsic equilibrium contact angles $\theta_1$ and $\theta_2$ are chosen, all of them leading to the same $\theta_{hete}$ in equation (\ref{eq:CB}). The critical Bond number is zero for the homogeneous case, while it is different from zero for the heterogeneous case.\label{fig:changeBo}}
\end{figure}

The top panel of Fig.~\ref{fig:changeBo} shows the position of the front contact point as a function of time for different \textcolor{black}{$\rm{Bo}$}. Increasing \textcolor{black}{$\rm{Bo}$}, we see that the net separation of time scales, characterizing the pinning of the drop and the jump forward, is progressively disappearing. The capillary number computed from the mean velocity of the drop is displayed as a function of \textcolor{black}{$\rm{Bo}$} in the bottom panel of Fig.~\ref{fig:changeBo}. We have chosen various wettabilities, producing the same Cassie-Baxter angle in Eq. (\ref{eq:CB}). At variance with the experimental data of Fig.~\ref{U_Ca_Bo}, sliding on homogeneous surfaces in numerical simulations is by construction not affected by the hysteresis. Therefore, from Fig.~\ref{fig:changeBo} we can appreciate the effect of the pattern in introducing a critical Bond number for the onset of motion, representing the increase of the static energetic barrier that must be overwhelmed by gravity before the drop starts to move. The slope \textcolor{black}{$\Delta \rm{Ca}/\Delta \rm{Bo}$} is basically unchanged if we keep fixed the effective contact angle provided by the Cassie-Baxter equation (\ref{eq:CB}), at least for \textcolor{black}{$\rm{Bo}$} reasonably larger than \textcolor{black}{$\rm{Bo_c}$}~\cite{Herdeetal12}.  This can be understood in terms of a simple qualitative argument allowing us to identify an effective angle, parametrizing the effective (average) dissipation at the contact line. For the homogeneous surface, viscous dissipation develops at the contact line and counterbalance the work done by the external (gravity) force on the drop. The viscous dissipation is parametrized by the dynamic angle $\theta_d$, which is close to the equilibrium angle $\theta_{eq}$ for small \textcolor{black}{Ca} and small hysteresis (see section~\ref{sec:1a}). The stationary wedge is therefore identified by the angle whose cosine projects the liquid-gas surface tension ($\gamma_{LG}$) to balance the difference between the solid-gas and solid-liquid surface tensions ($\gamma_{SG}-\gamma_{SL}$), i.e. Young equation $\gamma_{LG} \cos \theta_{eq} \approx \gamma_{SG}-\gamma_{SL}$.  In the heterogeneous case, when we seek the angle whose cosine projects the liquid-gas surface tension to balance the difference between the solid-gas and solid-liquid surface tensions  averaged over the period, we end up with the Cassie-Baxter prediction (\ref{eq:CB}).

\begin{figure}[!tbp]
\includegraphics[scale=0.3]{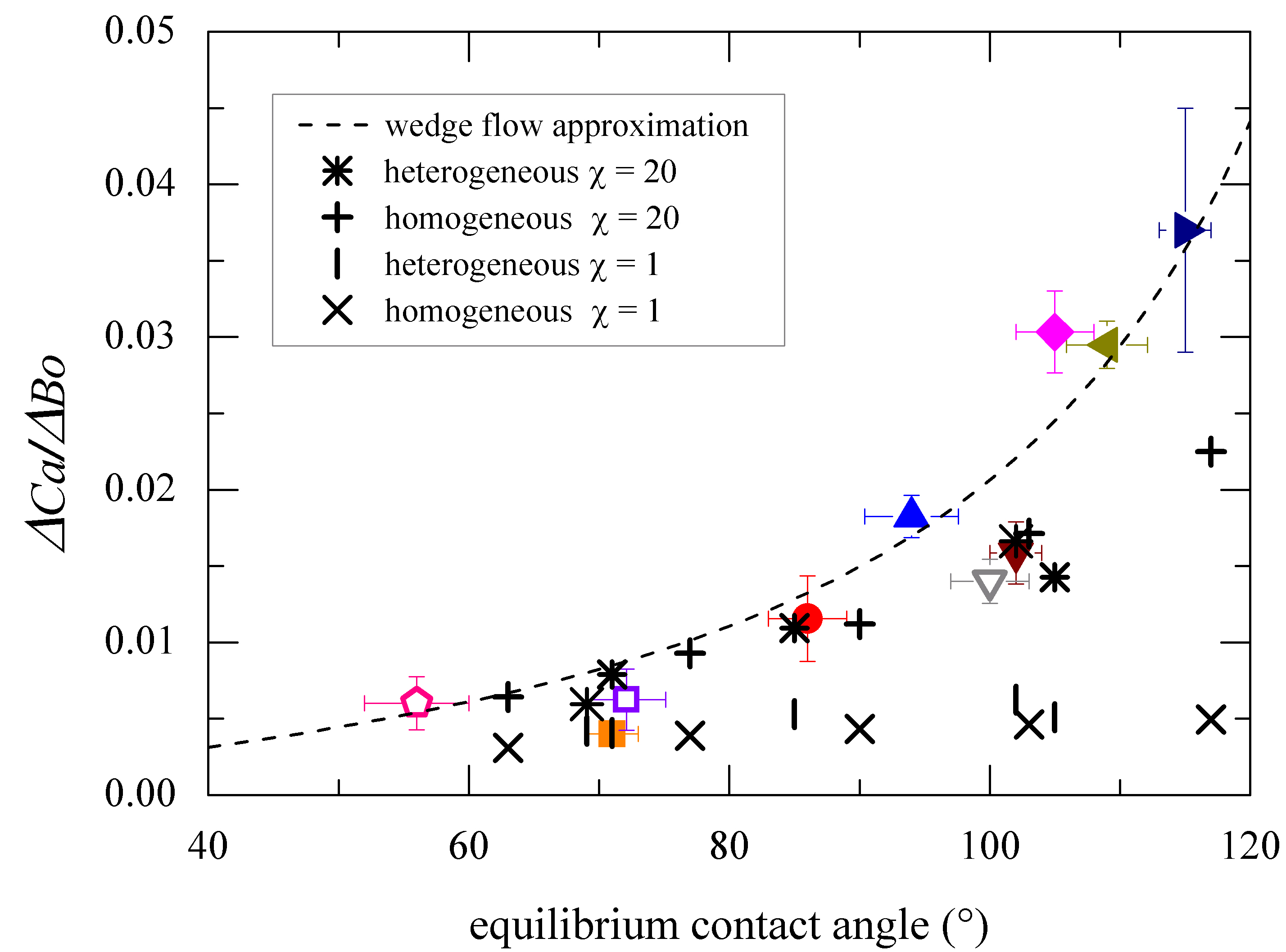}
\caption{(Color online) Slope of the \textcolor{black}{Ca} {\it vs.} \textcolor{black}{$\rm{Bo}$} curve as a function of the equilibrium contact angle. Open (filled) symbols refer to the heterogeneous (homogeneous) experimental values analyzed in Fig.~\ref{U_Ca_Bo}. Numerical simulations are performed with viscous ratios $\chi=1$ and $\chi=20$ between the drop phase and the outer phase for both the homogeneous and the heterogeneous cases. The dashed line is the scaling law predicted by (\ref{eq:scaling}), calculated for small drops sliding down homogeneous surfaces with a wedge dissipation as the dominant dissipative contribution~\cite{HuHScriven71,prl13, Kimetal02}. \label{fig:finale}}
\end{figure}

To check the validity of this argument against a change in the viscous ratio between the inner and outer drop regions, as well as a change in the fractions $f_1$ and $f_2$, we conducted a series of numerical simulations by changing the dynamic viscosity of the outer phase, exploring cases with $f_1=1/4,f_2=3/4$; $f_1=1/2,f_2=1/2$; $f_1=3/4,f_2=1/4$.  This offers the possibility to complement the results presented in Fig.~\ref{U_Ca_Bo} and extend the results presented in~\cite{prl13} which are limited to situations with $f_1=1/2,f_2=1/2$. In Fig.~\ref{fig:finale} we display the slope  \textcolor{black}{$\Delta \rm{Ca}/\Delta \rm{Bo}$}, including both the experimental data of Fig.~\ref{U_Ca_Bo} as well as the numerical results with two viscous ratios, $\chi=1$ and $\chi=20$. Similarly to what we have done for the experiments, we have performed numerical simulations for both homogeneous and heterogeneous samples. In all mesoscale approaches, as already noticed elsewhere~\cite{Kusumaatmajaeyal06}, the non-ideal interface is too wide (relative to the drop radius) with respect to the experiments. The resulting  contact line velocity is larger and the drop therefore moves too quickly in the simulations. This problem is accounted for by introducing a scaling factor, the same for all the numerical simulations. Such scaling factor is found to be of the order of the ratio $\log (\xi_{LB}/R_{LB})/ \log (\xi/R) \approx 0.2$, with $\xi$ the interface width (quantities without subscript refer to experimental values), as one would guess by looking at the solution of the laminar flow equations in a wedge~\cite{HuHScriven71,Kimetal02}. The numerical results with $\chi=1$ do not show any appreciable variation of the slope \textcolor{black}{$\Delta \rm{Ca}/\Delta \rm{Bo}$} with the equilibrium contact angle, indicating that the dissipation is unchanged at changing the equilibrium contact angle. For a drop sliding down a homogeneous surface with equilibrium contact angle $\theta_{eq}$, a flow develops in the outer wedge angled by an angle $\pi-\theta_{eq}$. Being the viscosity of the inner and outer phase the same, the dissipation for a system composed of a drop with equilibrium contact angle $\theta_{eq}$ is therefore the same as that of a drop with equilibrium contact angle $\pi-\theta_{eq}$. This symmetry in changing the outer fluid with the inner fluid is responsible for the independence of \textcolor{black}{$\Delta \rm{Ca}/\Delta \rm{Bo}$} on the contact angle. Repeating the simulations with the heterogeneous cases, we obtain the same value of \textcolor{black}{$\Delta \rm{Ca}/\Delta \rm{Bo}$}, \textcolor{black}{witnessing that the average dissipation for the patterned surfaces grows in a similar way at increasing the Bond number}. To observe a variation of the slope with respect to a change in the equilibrium contact angle, we need to change the viscous ratio $\chi$. Numerical results are shown for the case $\chi=20$: the change in the slope that we achieve is not as large as the one that we get in the experiments, and the reason is probably because such viscous ratio is still smaller than the experimental values. Unfortunately, numerical simulations with very large $\chi$ are quite unstable and technical improvements are needed to cure such numerical instabilities. At very large viscous ratio the dependence of the slope \textcolor{black}{$\Delta \rm{Ca}/\Delta \rm{Bo}$} as a function of the equilibrium contact angle $\theta_{eq}$ can be described by the scaling law Eq.~(\ref{eq:scaling}) with $c(\theta_{eq})$ calculated through the so called `wedge flow approximation'~\cite{HuHScriven71,Kimetal02,prl13}: such scaling law is reported for comparison with the experimental and numerical data. \textcolor{black}{One has also to note that the usual rescaling factor used in the numerical simulation is intimately connected to the idea that viscous dissipation is dominated by contact line dissipation. A recent detailed numerical study of dissipation loss inside sliding drops~\cite{Moradi} shows non negligible contributions from the region below the drop's center of mass. This leads to a refined scaling-law for the droplet velocity as a function of $\rm{Bo}$, which is different from the traditional scaling \cite{HuHScriven71}. This can also be a source of discrepancy between the numerical results and the experiments for large contact angles}. Here we recall that the scaling of \textcolor{black}{$\Delta \rm{Ca}/\Delta \rm{Bo}$} encodes the general feature that smaller contact angles are associated with higher viscous dissipation (see Section~\ref{sec:1}). Overall, the numerical simulations provide evidence that the slope \textcolor{black}{$\Delta \rm{Ca}/\Delta \rm{Bo}$} is well parametrized by the equilibrium contact angle, either homogeneous or heterogeneous, even in situations where the outer phase has a non negligible viscosity with respect to the drop phase.

\section{Conclusions}\label{sec:3}

We have characterized both experimentally and numerically the motion of drops sliding across alternating stripes having a large wettability contrast. For Bond numbers close to a critical Bond number, these drops undergo a characteristic non linear stick-slip motion whose average speed can easily be an order of magnitude smaller than that measured on a homogeneous surface having the same equilibrium contact angle. The slow down is the result of the pinning-depinning transition of the contact line which causes energy dissipation to be localized in time and large part of the driving energy to be stored in the periodic deformations of the contact line when crossing the stripes. We have quantified the change of dissipation inside the drop as a function of the increasing Bond number, by comparing the motion of the drops on heterogeneous patterns with those on homogeneous substrates: the main effects of the heterogeneous patterning can be readsorbed in a renormalized value of the critical Bond number, representing the increase of the static energetic barrier that must be overcome by gravity before the drop starts to move. Our findings suggest workable strategies to passively control the motion of drops by a suitable tailoring of the chemical pattern. It is also worth underscoring the essential role played by numerical simulations, which offer great flexibility in investigating a variety of load conditions and performing local measurements of  capillary, viscous and body forces, otherwise  impossible to obtain by experimental means. This would provide invaluable insights in the engineering of chemical patterns in open microfluidic devices. \\

\begin{acknowledgments}
We are particularly grateful to M. Brinkmann and C. Semprebon for useful discussions and to G. Dalle Rive for kind support in the sample preparation. We kindly acknowledge funding from the European Research Council under the Europeans Community's Seventh Framework Programme (FP7/2007-2013) / ERC Grant Agreement  N. 279004, the University of Padova, Italy (PRAT 2011 `MINET' and PRAT 2009 N. CPDA092517/09) and Fondazione Cariparo of Padova (Excellence Project call-2011, IOM-LiNbO).
\end{acknowledgments}

\section{Appendix}\label{sec:appendix}

The LB equation evolves in time the discretized probability density function $f_{\zeta i}({\bm{r}},t)$ to find at position ${\bm{r}}$ and time $t$ a fluid particle of component $\zeta=A,B$ with velocity  ${\bm{c}}_i$ according to the LB updating scheme
\begin{equation}\label{EQ:LBapp}
f_{\zeta i} ({\bm{r}} + \tau {\bm{c}}_i , t + \tau ) = f^{*}_{\zeta i} ({\bm{r}},t)= f_{\zeta i} ({\bm{r}},t) + \Delta_{\zeta i} +\Delta^{g}_{\zeta i}
\end{equation}
with the time step $\tau$ set to a unitary value. The (linear) collisional operator expresses the relaxation of the probability distribution function towards the local equilibrium $f^{(eq)}_{\zeta j}$ (the $*$ in (\ref{EQ:LBapp}) indicates the post-collisional probability density)
\be\label{eq:collis}
\Delta_{\zeta i}=\sum_{j} {\cal L}_{ij}(f_{\zeta j}-f^{(eq)}_{\zeta j})
\ee
where the expression for the equilibrium distribution is a result of the projection onto the lower order Hermite polynomials~\cite{Dunweg,DHumieres02} and the weights $w_i$ are {\it a priori} known through the choice of the quadrature
\be\label{feq}
f_{\zeta i}^{(eq)}=w_i \rho_{\zeta} \left[1+\frac{{\bm{u}} \cdot {\bm{c}}_i}{c_s^2}+\frac{{\bm{u}}{\bm{u}}:({\bm{c}}_i{\bm{c}}_i-{\bm I})}{2 c_s^4} \right]
\ee
\begin{equation}\label{weights}
w_i=
\begin{cases}
1/3 & i=0\\
1/18 & i=1\ldots6\\
1/36 & i=7\ldots18
\end{cases},
\end{equation}
where $c_s$ is the isothermal speed of sound (a constant in the model) and ${\bm{u}}$ is the fluid velocity. Our implementation features a D3Q19 model with 19 velocities
\begin{equation}\label{velo}
{\bm{c}}_i=
\begin{cases}
(0,0,0) & i=0\\
(\pm 1,0,0), (0,\pm 1,0), (0,0,\pm 1) & i=1\ldots6\\
(\pm 1,\pm 1,0), (\pm 1,0,\pm 1), (0,\pm 1,\pm 1)  & i=7\ldots18
\end{cases}.
\end{equation}
The operator ${\cal L}_{ij}$ in equation (\ref{eq:collis}) is the same for both components (this choice is appropriate when we  describe a symmetric binary mixture) and is constructed to have a diagonal representation in the so-called {\it mode space}:  the basis vectors ${\bm{e}}_{k}$ ($k=0,...,18$) of mode space are constructed by orthogonalizing polynomials of the dimensionless velocity vectors~\cite{Dunweg,DHumieres02}. The basis vectors are used to calculate a complete set of moments, the so-called modes $m_{\zeta k}=\sum_i {\bm{e}}_{ki} f_{\zeta i}$ ($k=0,...,18$). The lowest order modes are associated with the hydrodynamic variables. In particular, the zero-th order momenta give the densities for both components
\begin{equation}
\rho_{\zeta}=m_{\zeta 0}=\sum_{i} f_{\zeta i},
\end{equation}
with the total density given by $\rho=\sum_{\zeta}m_{\zeta 0} =\sum_{\zeta}\rho_{\zeta}$. The next three momenta $\tilde{\bm{m}}_{\zeta}=(m_{\zeta 1}, m_{\zeta 2}, m_{\zeta 3})$, when properly summed over all the components, are related to the velocity of the mixture
\begin{equation}
{\bm{u}} \equiv \frac{1}{\rho}\sum_{\zeta} \tilde{\bm{m}}_{\zeta}   +\frac{1}{2 \rho}\tau {\bm{g}} = \frac{1}{\rho}\sum_{\zeta} \sum_i f_{\zeta i} {\bm{c}}_{i}+\frac{1}{2 \rho}\tau {\bm{g}}.
\end{equation}
The other modes are the bulk and the shear modes (associated with the viscous stress tensor), and four groups of kinetic modes which do not emerge at the hydrodynamical level~\cite{Dunweg}. Since the operator ${\cal L}_{ij}$ is  diagonal in mode space, the collisional term describes a linear relaxation of the non-equilibrium modes
\be\label{MODES}
m^{*}_{\zeta k}=(1+\lambda_k)m_{\zeta k}+m_{\zeta k}^{g}
\ee
where the relaxation frequencies $-\lambda_k$ (i.e. the eigenvalues of $-{\cal L}_{ij}$) are related to the transport coefficients of the modes. The term $m_{\zeta k}^{g}$ is related to the $k$-th moment of the forcing source $\Delta_{\zeta i}^{g}$ associated with a forcing term with density ${\bm{g}}_{\zeta}({\bm{r}}, t)$.  While the forces have no effect on the mass density, they transfer an amount ${\bm{g}}_{\zeta}\tau$ of total momentum to the fluid in one time step. The forcing term is determined in such a way that the hydrodynamical equations (\ref{eq:2}-\ref{eq:3}) are recovered, and can be written as \cite{Guo}
\begin{equation}
\Delta_{\zeta i}^{g}=\frac{w_i \tau}{c_s^2} \left(\frac{2+\lambda_M}{2}\right) {\bm{g}}_{\zeta} \cdot {\bm{c}}_i\\ +\frac{w_i \tau}{c_s^2} \left[\frac{1}{2c_s^2} {\bm{G}} : ({\bm{c}}_i {\bm{c}}_i-c_s^2 {\bm I} ) \right],
\end{equation}
where the components of tensor ${\bm G}$ are defined as
\begin{equation}
G_{\alpha \beta}=\frac{2+\lambda_s}{2}\left(u_{\alpha} g_{\zeta \beta}+g_{\zeta \alpha} u_{\beta}-\frac{2}{3} u_{\gamma} g_{\zeta \gamma} \delta_{\alpha \beta} \right)+\frac{2+\lambda_b}{3} u_{\gamma} g_{\zeta \gamma} \delta_{\alpha \beta}.
\end{equation}
Using the LB model we are able to reproduce the continuity equations and the Navier Stokes equations for both densities (repeated indexes are meant summed upon)~\cite{SegaSbragaglia13}
\be\label{eq:2}
\frac{\de}{\de t} \rho_{\zeta}+\frac{\partial}{\partial r_{\beta}} (\rho_{\zeta} u_{\beta}) = \partial_{\beta} D_{\zeta \beta},
\ee
\be\label{eq:3}
\rho\left[\frac{\partial u_{\al}}{\partial t}+u_{\beta} \frac{\partial u_{\al}}{\partial r_{\beta}}  \right]=-\frac{\partial p}{\partial r_{\alpha}}+\frac{\partial \sigma_{\alpha \beta}}{\partial r_{\beta}} + {g}_{\zeta \alpha}.
\ee
In the above equations, $\rho=\sum_{\zeta}\rho_\zeta$ is the total density and $p=\sum_{\zeta} p_{\zeta}=\sum_{\zeta} c_s^2 \rho_{\zeta}$ is the internal pressure of the mixture.  The $\alpha$-th projection of the velocity is denoted with ${u}_{\alpha}$. The term $\sum_{\zeta} {g}_{\zeta \alpha}$ refers to all the contributions coming from internal and external forces. As for the internal forces, we will use the ``Shan-Chen'' model~\cite{SC} for multicomponent mixtures
\begin{equation}\label{eq:SCforce}
{g}_{\zeta \alpha}({\bm{r}}) =  - \rho_{\zeta}({\bm{r}}) \sum_{i} \sum_{\zeta'\neq \zeta} w_i  g_{A B} \rho_{\zeta^{\prime}} ({\bm{r}}+\tau \bm{c}_i) {c}_{i \alpha} \hspace{.2in} \zeta,\zeta^{\prime}=A,B
\end{equation}
where $g_{A B}$ is a function that regulates the interactions between different pairs of components. The sum in equation (\ref{eq:SCforce}) extends over a set of interaction links coinciding with those of the LB dynamics (see equation (\ref{velo})). When the coupling strength parameter $g_{A B}$ is sufficiently large, demixing occurs and the model can describe stable interfaces with a surface tension. The effect of the internal forces can be recast into the gradient of the pressure tensor $P^{(int)}_{\alpha \beta}$~\cite{SbragagliaBelardinelli}, thus modifying the internal pressure of the model, i.e. $P_{\alpha \beta} = p \, {\delta}_{\alpha \beta}+P^{(int)}_{\alpha \beta}$.  The thermodynamic properties of the drop are input via such a pressure tensor: this accounts for the surface tension at the interface between the two fluids, as well as the capillary forces at the contact line via a suitable imposition of wetting boundary conditions for the densities at the wall. The diffusion current ${\bm{D}}_{\zeta}$ and the viscous stress tensor ${\bm{\sigma}}$ in equations (\ref{eq:2}-\ref{eq:3}) are given by
\begin{equation}
D_{\zeta \alpha}=\mu \left[\left(\frac{\de p_{\zeta}}{\de r_{\alpha}}-\frac{\rho_{\zeta}}{\rho} \frac{\de p}{\de r_{\alpha}}\right)-\left(g_{\zeta \alpha}-\frac{\rho_{\zeta}}{\rho}g_{\alpha}\right) \right], \hspace{.4in} \sigma_{\alpha \beta} =\eta_s{\left(\frac{\de u_{\beta}}{\de r_{\alpha}} +\frac{\de u_{\alpha}}{\de r_{\beta}} -\frac{2}{3}\frac{\de u_{\gamma}}{\de r_{\gamma}} \delta_{\alpha \beta}\right)}+\eta_b \frac{\de u_{\gamma}}{\de r_{\gamma}} \delta_{\alpha \beta}.
\label{eq:comp_Pi}
\end{equation}
The relaxation times of the momentum ($\lambda_M$), bulk ($\lambda_b$) and shear ($\lambda_s$) modes in (\ref{eq:collis}) are related to the transport coefficients of hydrodynamics as
\begin{equation}\label{TRANSPORTCOEFF}
\mu=-\tau \left(\frac{1}{\lambda_M}+\frac{1}{2} \right) \hspace{.2in} \eta_s=-\rho c_s^2 \tau \left(\frac{1}{\lambda_s}+\frac{1}{2} \right) \hspace{.2in} \eta_b=-\frac{2}{3}\rho c_s^2 \tau \left(\frac{1}{\lambda_b}+\frac{1}{2} \right)
\end{equation}
where $\mu$ is the mobility and $\eta_b$, $\eta_s$ the bulk and shear viscosities respectively. We introduce the effect of gravity in the Navier-Stokes equation with a body force density, $\rho _{A}g \sin \alpha$, applied to the $A$ phase along the $x$-direction. For the numerical simulations presented we have used $g_{A B}=1.5$ lbu (LB units) in (\ref{eq:SCforce}) corresponding to a surface tension $\gamma_{LG}=0.2$ lbu and associated bulk densities $\rho_A=2.3$ lbu and $\rho_B=0.06$ lbu in the $A$-rich region. The relaxation frequencies in (\ref{TRANSPORTCOEFF}) are such that $\lambda_M=\lambda_s=\lambda_b=-1.0$ lbu, corresponding to a viscous ratio $\chi=\eta_{in}/\eta_{out}=1$, where $\eta_{in}$, $\eta_{out}$ are the dynamic viscosities inside (inner viscosity) and outside (outer viscosity) the drop, respectively. The cases with $\chi \neq 1$ are obtained by letting $\lambda_s$ depend on the component $\zeta$, thus allowing to model an inner dynamic viscosity larger than the outer one.


\end{document}